%% file: paper.tex
\documentclass[twocolumn,showpacs,aps,prd,floatfix]{revtex4}

\usepackage{amsmath} 
\usepackage{slashbox} 
\usepackage{multirow} 
\usepackage{graphicx} 

\def\Babar{\slshape B\kern-0.1em{\footnotesize A}\kern-0.1em B\kern-0.10em{\footnotesize A\kern-0.20em R}} 
\def\babar{\slshape B\kern-0.1em{\scriptsize A}\kern-0.1em B\kern-0.10em{\scriptsize A\kern-0.20em R}} 
\def\Bbar{\kern 0.18em\overline{\kern -0.18em B}{}} 
\def\BbarSubscr{\kern 0.25em\overline{\kern -0.18em B}{}} 
\def\Dbar{\kern 0.20em\overline{\kern -0.20em D}{}} 
\def\MES{M_{\rm ES}} 
\def\DELTAE{\Delta E} 
\def\gev{\mathrm{\,Ge\kern -0.1em V}} 
\def\mev{\mathrm{\,Me\kern -0.1em V}} 
\def\gevc{\mathrm{\,Ge\kern -0.1em V\!/}c} 
\def\gevcc{\mathrm{\,Ge\kern -0.1em V\!/}c^2} 
\def\mevcc{\mathrm{\,Me\kern -0.1em V\!/}c^2} 
\def\gevccsq{\mathrm{\,Ge\kern -0.1em V^2\!/}c^4} 
\def\ubar{\kern 0.10em \overline{\kern-0.10em u}\kern 0.05em{}} 
\def\dbar{\kern 0.15em \overline{\kern-0.15em d}\kern 0.05em{}} 
\def\sbar{\kern 0.10em \overline{\kern-0.10em s}\kern 0.05em{}} 
\def\cbar{\kern 0.10em \overline{\kern-0.10em c}\kern 0.05em{}} 
\def\bbar{\kern 0.10em \overline{\kern-0.10em b}\kern 0.05em{}} 
\def\nbar{\kern 0.10em \overline{\kern-0.10em n}\kern 0.05em{}} 
\def\pbar{\kern 0.10em \overline{\kern-0.10em p}\kern 0.05em{}} 
\def\Nbar{\kern 0.25em \overline{\kern-0.25em N}\kern 0.05em{}} 
\def\dbline{\noalign{\vskip 0.10truecm\hrule\vskip 0.05truecm\hrule\vskip 0.10truecm}} 
\def\sgline{\noalign{\vskip 0.10truecm\hrule\vskip 0.10truecm}} 
\def\bentarrow{\mbox{$\:\raisebox{1.4ex}{\rlap{$\big\vert$}}\!\rightarrow\,$}} 
\def\bang{\!\!\!}
\def\Bang{\!\!\!\!\!}
\def\BANG{\!\!\!\!\!\!\!}
\long\def\inst#1{\par\nobreak\kern 4pt\nobreak {\it #1}\par\vskip 10pt plus 3pt minus 3pt} 

\begin{document} 

\widetext 

\begin{flushleft}
    SLAC-PUB-14763 \\
    {\Babar}-PUB-10/019
\end{flushleft}

\title{
    \boldmath Observation and study of the baryonic
    $\protect{B}$-meson decays
    $\protect{B}{\rightarrow}{D^{(\!\ast\!)}}{p\pbar}(\!\pi\!)(\!\pi\!)$
}

\input authors_jun2010_bad2210.tex

\begin{abstract} 
\noindent  
    We present results for $B$-meson decay modes involving a charm
    meson, protons, and pions using $455{\times}10^6$ $B\Bbar$ pairs
    recorded by the {\babar} detector at the SLAC PEP-II
    asymmetric-energy ${e^+e^-}$ collider.  The branching fractions
    are measured for the following ten decays:
    ${\Bbar^0}{\rightarrow}{D^0}{p\pbar}$,
    ${\Bbar^0}{\rightarrow}{D^{\ast0}}{p\pbar}$,
    ${\Bbar^0}{\rightarrow}{D^+}{p\pbar}{\pi^-}$,
    ${\Bbar^0}{\rightarrow}{D^{\ast+}}{p\pbar}{\pi^-}$,
    ${B^-}{\rightarrow}{D^0}{p\pbar}{\pi^-}$,
    ${B^-}{\rightarrow}{D^{\ast0}}{p\pbar}{\pi^-}$,
    ${\Bbar^0}{\rightarrow}{D^0}{p\pbar}{\pi^-}{\pi^+}$,
    ${\Bbar^0}{\rightarrow}{D^{\ast0}}{p\pbar}{\pi^-}{\pi^+}$,
    ${B^-}{\rightarrow}{D^+}{p\pbar}{\pi^-}{\pi^-}$, and
    ${B^-}{\rightarrow}{D^{\ast+}}{p\pbar}{\pi^-}{\pi^-}$.
    The four $B^-$ and the two five-body $\Bbar^0$ modes are observed
    for the first time.  The four-body modes are enhanced compared to
    the three- and the five-body modes.  In the three-body modes, the
    $M({p\pbar})$ and $M({D^{(\!\ast\!)0}}{p})$ invariant mass
    distributions show enhancements near threshold values.  In the
    four-body mode ${\Bbar^0}{\rightarrow}{D^+}{p\pbar}{\pi^-}$, the
    $M(p{\pi^-})$ distribution shows a narrow structure of unknown
    origin near $1.5\gevcc$.  The distributions for the five-body
    modes, in contrast to the others, are similar to the expectations
    from uniform phase-space predictions. 
\end{abstract} 
\maketitle 

\section{INTRODUCTION} 
\label{sec:introduction} 

\noindent
$B$-meson decays to final states with baryons have been explored much
less systematically than decays to meson-only final states.  The first
exclusively reconstructed decay modes were the CLEO observations of
$B{\rightarrow}\Lambda_c^+\,\pbar\pi$ and
$B{\rightarrow}\Lambda_c^+\,\pbar\pi\pi$ \cite{Fu:1996qt} and, later,
of ${\Bbar^0}{\rightarrow}{D^{\ast+}}{p\pbar}{\pi^-}$ and
${B^0}{\rightarrow}{D^{\ast-}}{p}\nbar$ \cite{Anderson:2000tz}.  These
measurements supported the prediction \cite{Dunietz:1998uz} that the
final states with $\Lambda_c$ baryons are not the only sizable
contributions to the baryonic $B$-meson decay rate, and that the
charm-meson modes of the form
${B}{\rightarrow}{D^{(\!\ast\!)}}\!{N}\!\Nbar^{\prime}\!{+}\textit{anything}$,
where the $N^{(\prime)}$ represent nucleon states, are also
significant.  Previous measurements show a trend that the branching
fractions increase with the number of final-state particles.  The
branching fractions for the four-body modes
${\Bbar^0}{\rightarrow}{D^{(\!\ast\!)+}}{p\pbar}{\pi^-}$
\cite{Anderson:2000tz, Aubert:2006qx} are approximately four times
larger than those for the three-body modes
${\Bbar^0}{\rightarrow}{D^{(\!\ast\!)0}}{p\pbar}$ \cite{Abe:2002tw},
which, in turn, is approximately five times larger than those for the
two-body modes $\Bbar^0{\rightarrow}\Lambda_c^+\,\pbar$
\cite{Aubert:2008if}.

We expand the scope of baryonic $B$-decay studies with measurements of
the branching fractions and the kinematic distributions of the
following ten modes \cite{fn:ckm, fn:charge}:

\begin{center} 
\begin{tabular}{clll} 
\multicolumn{1}{l}{Three-body}
    &${\Bbar^0}{\rightarrow}{D^0}{p\pbar}$ 
    &\!and\!  &${\Bbar^0}{\rightarrow}{D^{\ast0}}{p\pbar}$,\\ 
\multicolumn{1}{l}{Four-body}
    &${\Bbar^0}{\rightarrow}{D^+}{p\pbar}{\pi^-}$ 
    &\!and\!  &${\Bbar^0}{\rightarrow}{D^{\ast+}}{p\pbar}{\pi^-}$,\\ 
    $''$      &${B^-}{\rightarrow}{D^0}{p\pbar}{\pi^-}$ 
    &\!and\!  &${B^-}{\rightarrow}{D^{\ast0}}{p\pbar}{\pi^-}$,\\ 
\multicolumn{1}{l}{Five-body}
    &${\Bbar^0}{\rightarrow}{D^0}{p\pbar}{\pi^-}{\pi^+}$ 
    &\!and\!  &${\Bbar^0}{\rightarrow}{D^{\ast0}}{p\pbar}{\pi^-}{\pi^+}$,\\ 
    $''$      &${B^-}{\rightarrow}{D^+}{p\pbar}{\pi^-}{\pi^-}$ 
    &\!and\!  &${B^-}{\rightarrow}{D^{\ast+}}{p\pbar}{\pi^-}{\pi^-}$.
\end{tabular} 
\end{center} 

\noindent 
Six of the modes---the four $B^-$ and the two five-body $B^0$
modes---are observed for the first time.

We reconstruct the modes through twenty-six decay chains consisting of
all-hadronic final states (the list is given later with the results in
Table~\ref{tab:bf_chain}), e.\,g., 

\begin{center} 
\begin{tabular}{rcll} 
    ${B^+}$ & \!$\rightarrow$\! 
    & \multicolumn{2}{l}{${D^{\ast-}}{p\pbar}{\pi^+}{\pi^+}$}\\ 
    & & \bentarrow & ${\Dbar^0}{\pi^-}$\\ 
    & & & \bentarrow ${K^+}{\pi^-}{\pi^+}{\pi^-}$. 
\end{tabular} 
\end{center} 

\noindent
A $D^0$ meson, as in the above example, is produced in eight of the
$B$ modes and a $D^+$ is produced in the remaining two.  The
${D^0}$-meson candidates are reconstructed through decays to
${K^-}{\pi^+}$, ${K^-}{\pi^+}{\pi^0}$, and
${K^-}{\pi^+}{\pi^-}{\pi^+}$; and the ${D^+}$ to
${K^-}{\pi^+}{\pi^+}$.  The ${D^{\ast0}}$-meson candidates are
reconstructed through decays to ${D^0}{\pi^0}$ and the ${D^{\ast+}}$
as ${D^0}{\pi^+}$.

\begin{figure}[tbp!] 
\centering 
\includegraphics[width=0.35\textwidth]{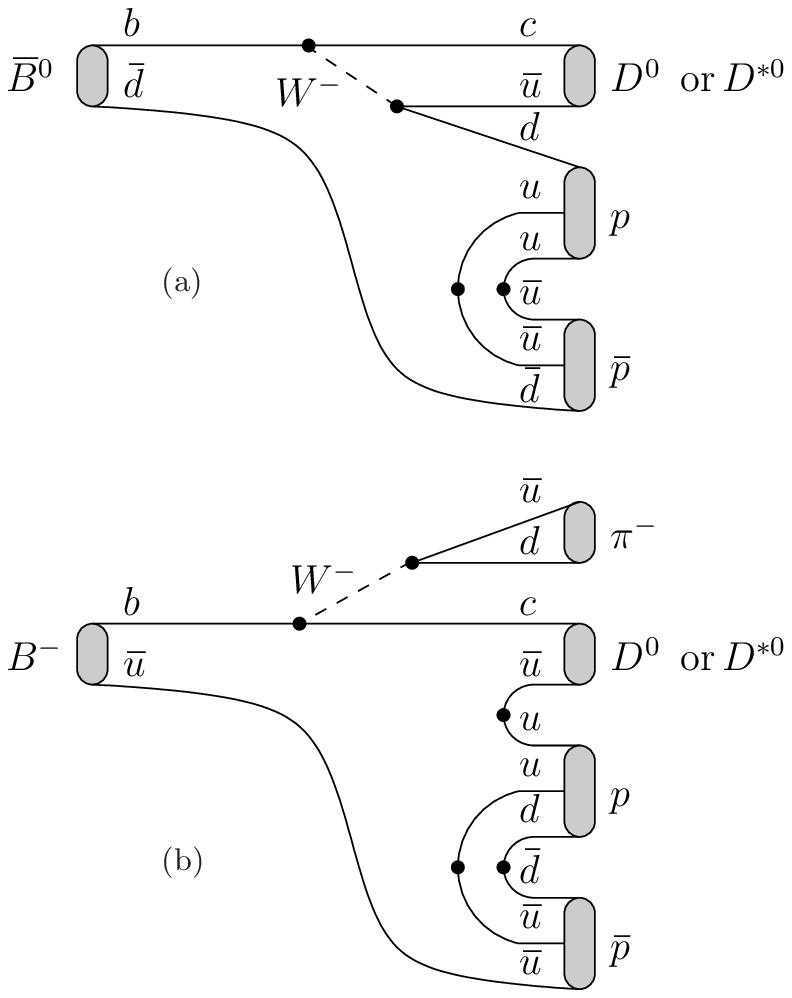} 
\caption{ 
    Typical quark-line diagrams representing (a) 
    ${\Bbar^0}{\rightarrow}{D^{(\!\ast\!)0}}{p\pbar}$ and (b) 
    ${B^-}{\rightarrow}{D^{(\!\ast\!)0}}{p\pbar}{\pi^-}$ modes.
    The gluon lines are omitted. 
    }\label{fig:diagram} 
\end{figure}

Typical quark-line diagrams for the three- and four-body modes with a
$D^{(\ast)0}$ meson are shown in Fig.~\ref{fig:diagram}.  The
three-body modes involve internal emissions of the ${W^-}$ boson,
whereas the four- and five-body modes involve internal and external
emission diagrams.

Baryonic $B$ decays have a distinctive phenomenology whose features
contrast with the patterns observed in meson-only final states.
Experimentally, the overall rate enhancement of multi-body decays and
the low-mass enhancement in the baryon-antibaryon subsystem are
observed \cite{
Lee:2004mg,%
Wang:2005fc,%
Medvedeva:2007zz,%
Wei:2007fg,%
Chen:2008jy,%
Aubert:2005gw%
}.  Theoretically, these modes are used to investigate a wide range of
topics \cite{% 
Hou:2000bz,% 
Chua:2002,% 
Cheng:2002,% 
Bigi:2002,% 
Chang:2002,% 
Cheng:2001tr,% 
Cheng:2003fq,% 
Luo:2003pv,% 
Hsiao:2009,% 
Suzuki:2005iq,%
Hong:2009%
}. Among them are the predictions of the relative branching fractions,
the decay dynamics, and the hypotheses involving exotic QCD phenomena,
such as \text{tetra-}, \text{penta-}, or septa-quark bound states.  In
particular, there have been discussions of ${p\pbar}$ peaks near
threshold values and penta-quark intermediate resonance decays
$\Theta_c{\rightarrow}{D^{(\!\ast\!)+}}\pbar$ with respect to our
modes \cite{%
Rosner:2003bm,%
Datta:2003iy,%
Kerbikov:2004gs,%
Chang:2005%
}.

The paper is organized as follows: Section~\ref{sec:babar} describes
the data sample and the {\Babar} detector.  Section~\ref{sec:analysis}
presents the analysis method, introducing the key variables $\MES$ and
$\DELTAE$.  Section~\ref{sec:bf} shows the fits to the joint
$\MES$-$\DELTAE$ distributions.  The fit yields and the corresponding
branching fractions are given.  Section \ref{sec:syst} discusses the
systematic uncertainties.  Section~\ref{sec:dynamics} presents the
kinematic distributions. For the three-body modes, the Dalitz plots of
$M^2({D^{(\!\ast\!)0}}{p})$ vs.~$M^2({p\pbar})$ are given as well as
the invariant mass plots of the variables.  For the four- and
five-body modes, the two-body subsystem invariant mass plots are
given.  In the four-body modes, we investigate a narrow structure in
the $M(p{\pi^-})$ distribution near $1.5\gevcc$.
Section~\ref{sec:conclusions} states the conclusions.

\section{\boldmath {\Babar} DETECTOR AND DATA SAMPLE} 
\label{sec:babar} 

\noindent
We use a data sample with integrated luminosity of
$414\textrm{\,fb}^{-1}$ ($455{\times}10^6$ $B\Bbar$) recorded at the
center-of-mass energy $\sqrt{s}{=}10.58\gev$ with the {\Babar}
detector at the PEP-II $e^+e^-$ collider.  The $e^+$ and $e^-$ beams
circulate in the storage rings at energies of $3.1\gev$ and $9\gev$,
respectively.  The value of $\sqrt{s}$ corresponds to the
$\Upsilon{(4S)}$ mass, maximizing the cross section for
${e^+e^-}{\rightarrow}\,b\bbar{\rightarrow}\Upsilon{(4S)}{\rightarrow}B\Bbar$
events.  The $B\Bbar$ production accounts for approximately a quarter
of the total hadronic cross section; the continuum processes
${e^+e^-}{\rightarrow}\,u\ubar$, $d\dbar$, $s\sbar$, and $c\cbar$
constitute the rest.

The main components of the {\Babar} detector \cite{Aubert:2001tu} are
the tracking system, the Detector of Internally-Reflected Cherenkov
radiation (DIRC), the electromagnetic calorimeter, and the
instrumented flux return.

The two-part charged particle tracking system measures the momentum.
The silicon vertex tracker, with five layers of double-sided silicon
micro-strips, is closest to the interaction point.  The tracker is
followed by a wire drift chamber filled with a helium-isobutane
($80$:$20$) gas mixture, which was chosen to minimize multiple
scattering.  The superconducting coil creates a $1.5\,\textrm{T}$
solenoidal field.

The DIRC measures the opening angle of the Cherenkov light cone,
$\theta_\textrm{C}$, produced by a charged particle traversing one of
the 144 radiator bars of fused silica.  The light propagates in the
bar by total internal reflection and is projected onto an array of
photomultiplier tubes surrounding a water-filled box mounted at the
back end of the tracking system.  The DIRC's ability to distinguish
pions, kaons, and protons complements the energy loss measurements,
$\mathrm{d}E/\mathrm{d}x$, in the tracking volume.  

The calorimeter measures the energies and positions of electron-photon
showers with an array of 6580 finely-segmented Tl-doped CsI crystals.

The flux return is instrumented with a combination of resistive plate
chambers and limited streamer tubes for the detection of muons and
neutral hadrons.

\begin{figure}[tbp!] 
\centering 
\includegraphics[width=0.48\textwidth]{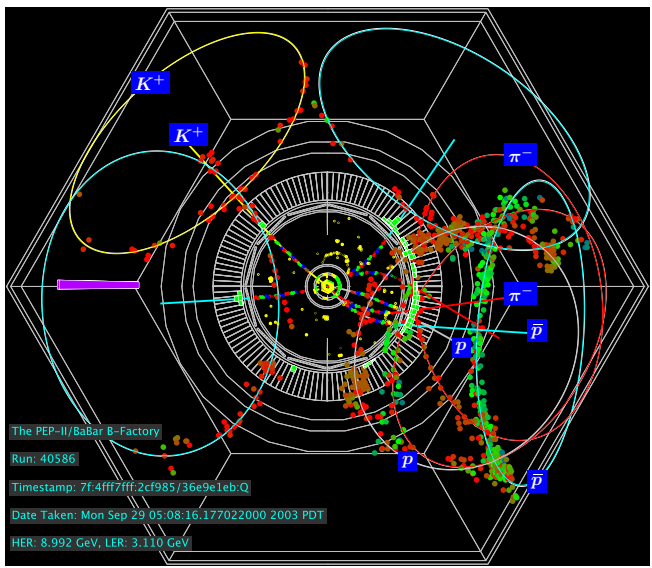} 
\caption{ 
    Event display for the candidate decay
    ${B^0}{\rightarrow}{\Dbar^0}{p\pbar}$,
    ${\Dbar^0}{\rightarrow}{K^+}{\pi^-}$. The labeled tracks in the
    tracking system and DIRC rings at the perimeter correspond to the
    particles in the reconstructed decay chain.  The remaining
    unlabeled tracks and rings are due to the decay of the other
    ${\Bbar^0}$ meson in the event.  The beam axis is perpendicular to
    the image.   
    }\label{fig:display}
 \end{figure}

A data event display is given in Fig.~\ref{fig:display} for the
candidate decay ${B^0}{\rightarrow}{\Dbar^0}{p\pbar}$,
${\Dbar^0}{\rightarrow}{K^+}{\pi^-}$.

\section{ANALYSIS METHOD} 
\label{sec:analysis} 

\noindent
This section describes the branching fraction measurement in four
parts. Section~\ref{sec:mcsamples} describes the Monte Carlo-simulated
event samples that are used to evaluate the performance of the method.
Section~\ref{sec:selection} lists the discriminating variables and
their requirements for the event selection.  Section~\ref{sec:data}
defines the $\MES$ and $\DELTAE$ variables and presents their
distributions for the newly observed modes.  Lastly,
Sec.~\ref{sec:fit} describes the fit to the $\MES$-$\DELTAE$
distribution used to extract the signal yield.

\subsection{Monte Carlo-simulated event samples} 
\label{sec:mcsamples} 

\noindent
Monte Carlo (MC) event samples are produced and used to evaluate the
analysis method.  Two types of samples---signal and generic---are
described below.

The particle decays are generated using a combination of
\textsc{Evtgen} \cite{Lange:2001uf} and \textsc{Jetset\,7.4}
\cite{Sjostrand:1993yb}.  The interactions of the decay products
traversing the detector are modeled by \textsc{Geant\,4}
\cite{Agostinelli:2002hh}.  The simulation takes into account varying
detector conditions and beam backgrounds during the data-taking
periods.

The signal MC sample is generated to characterize events with a $B$
meson that decays to one of the signal modes (the accompanying $\Bbar$
decays generically).  The typical size of $3{\times}10^5$ events per
decay chain is two orders of magnitude larger than the expected signal
in data.  

The generic MC sample is generated to characterize the entire data
sample.  The size is approximately twice that of the {\Babar} data
sample.  

\subsection{Event selection} 
\label{sec:selection} 

\noindent
The $e^+e^-$ events are filtered for a signal $B$-meson candidate
through the pre- and the final selections.

The pre-selection requires the presence of proton-antiproton pair and
a ${D^0}$- or a ${D^+}$-meson candidate (written as $D$ without a
charge designation) in one of the 26 decay chains listed in
Sec.~\ref{sec:introduction}.

Protons are identified with a likelihood-based algorithm using the
$\mathrm{d}E/\mathrm{d}x$ and the $\theta_\textrm{C}$ measurements as
described in Sec~\ref{sec:babar}.  For a $1.0\gevc$ proton in the lab
frame (typical of those produced in a signal mode), the selection
efficiency is $98\%$ and the kaon fake rate is $1\%$.

The $D$-meson candidates are selected using the invariant mass
\cite{fn:dmass}, $M(D)$, and a kaon identification algorithm similar
to that used for protons.  The $M(D)$ is required to be within seven
times its resolution around the PDG value \cite{Amsler:2008zzb}
(superseded later during final selection).  For a $0.9\gevc$ kaon in
the lab frame (typical of those produced in a signal mode), the
selection efficiency is $85\%$ and the pion fake rate is $2\%$.

For the ${D^0}{\rightarrow}{K^-}{\pi^+}{\pi^0}$ and
${D^{\ast0}}{\rightarrow}{D^0}{\pi^0}$ sub-decay modes, the
${\pi^0}{\rightarrow}\,\gamma\gamma$ candidates are formed from two
well-separated photons with $115{<}M(\gamma\gamma){<}150\mevcc$ or
from two unseparated photons by using the second moment of the
overlapping calorimeter energy deposits. 

The charged particles from the decay chain are required to have a
distance of closest approach to the beam spot of less than $1.5$\,cm.

The final selection requires the presence of a fully-reconstructed
signal $B$-meson candidate.  Requirements on the discriminating
variables described below are optimized by maximizing the signal
precision $z{=}S/\sqrt{S{+}B}$, where $S$ is the expected signal yield
using the signal MC sample and $B$ the expected background yield using
the generic MC sample.  The signal is normalized using the measured
branching fractions for the modes
${\Bbar^0}{\rightarrow}{D^{(\!\ast\!)0}}{p\pbar}$ and
${\Bbar^0}{\rightarrow}{D^{\ast+}}{p\pbar}{\pi^-}$
\cite{Anderson:2000tz,Abe:2002tw}; for the rest of the modes the
latter value is used.  The quantity $z$ is computed for each
discriminating variable for each decay chain.  For the variables with
a broad maximum in $z$, the cut values are chosen to be consistent
across similar modes. 

In order to select $D$-meson candidates, $M(D)$ is required to be
within $3\,\sigma_{M\!(\!D\!)}$ of the PDG value
\cite{Amsler:2008zzb}.  The resolutions $\sigma_{M\!(\!D\!)}$ for
${D^0}{\rightarrow}{K^-}{\pi^+}$, ${K^-}{\pi^+}{\pi^0}$,
${K^-}{\pi^+}{\pi^-}{\pi^+}$, and
${D^+}{\rightarrow}{K^-}{\pi^+}{\pi^+}$ are approximately $6$, $10$,
$5$, and $5\mevcc$, respectively.  For the modes involving
${D^0}{\rightarrow}{K^-}{\pi^+}{\pi^0}$ decays, the combinatoric
background events due to fake ${\pi^0}$ candidates are suppressed
using a model \cite{Frabetti:1994di} that parameterizes the amplitude
of the Dalitz plot distribution $M^2({K^-}{\pi^+})$
vs.~$M^2({\pi^+}{\pi^0})$.  The model accounts for the amplitudes and
the interferences of decays of ${K^{\ast0}}{\rightarrow}K^-\pi^+$,
${K^{\ast-}}{\rightarrow}K^-\pi^0$, and
$\rho^+{\rightarrow}\pi^+\pi^0$.  The normalized magnitude of the
decay amplitude is used to suppress the background events by requiring
the quantity to be greater than a value ranging from $1\%$ to $5\%$,
depending on the mode.

In order to select ${D^\ast}$-meson candidates, the ${D^\ast}$-$D$
mass difference, $\Delta{M}{=}M({D^0}\pi){-}M({D^0})$, is required to
be within $3\,\sigma_{\Delta\!M}$ of the PDG value
\cite{Amsler:2008zzb}.  The resolution $\sigma_{\Delta\!M}$ is
approximately $0.8\mevcc$ for both $D^{\ast0}{\rightarrow}D^0\pi^0$
and $D^{\ast+}{\rightarrow}D^0\pi^+$.  For the mode
${\Bbar^0}{\rightarrow}{D^0}{p\pbar}{\pi^-}{\pi^+}$, the requirement
of $\Delta{M}{>}160\mevcc$ excludes the contamination from
${\Bbar^0}{\rightarrow}{D^{\ast+}}{p\pbar}{\pi^-}$,
${D^{\ast+}}{\rightarrow}{D^0}{\pi^+}$ decays. 

In order to select $B$-meson candidates, a combination of daughter
particles in one of the signal modes is considered.  The momentum
vectors of the decay products are fit \cite{Hulsbergen:2005pu} while
constraining $M(D)$ to the PDG value \cite{Amsler:2008zzb}.  The
vertex fit $\chi^2$ probability for non-$B$ events peaks sharply at
zero; these events are suppressed by requiring the probability to be
greater than $0.1\%$.

Continuum backgrounds events are suppressed by using the angle
$\theta_\textrm{thrust}$ between the thrust axes \cite{Aubert:2001xs}
of the particles from the $B$-meson candidate and from the rest of the
event.  The continuum event distribution of
$|\cos\theta_\textrm{thrust}|$ peaks at unity while it is uniform for
$B\Bbar$ events, so the quantity is required to be less than a value
ranging from $0.8$ to $1$, depending on the mode.

After the selection, an average of $1.0$ to $1.7$ candidates per event
remains for each decay chain and is largest for those decay chains
with the largest particle multiplicity.  If more than one candidate is
present, we choose the one with the smallest value of 
\begin{equation} 
    \delta = 
    \frac{(M(D)-M(D)_{\footnotesize\rm PDG})^2}{(\sigma_{M\!(\!D\!)})^2} + 
    \frac{(\Delta{M}-\Delta{M}_{\footnotesize\rm PDG})^2}{(\sigma_{\Delta\!M})^2}, 
\end{equation} 

\noindent
where the PDG values \cite{Amsler:2008zzb} are labeled as such.  The
latter term in the sum is included only if a ${D^\ast}$ is present in
the decay chain.  If more than one candidate has the same $\delta$
value, we choose one randomly.

\subsection{\boldmath Definitions of $\protect\MES$ and $\protect\DELTAE$} 
\label{sec:data} 

\begin{figure*}[tbp!] 
\centering 
\includegraphics[width=1\textwidth]{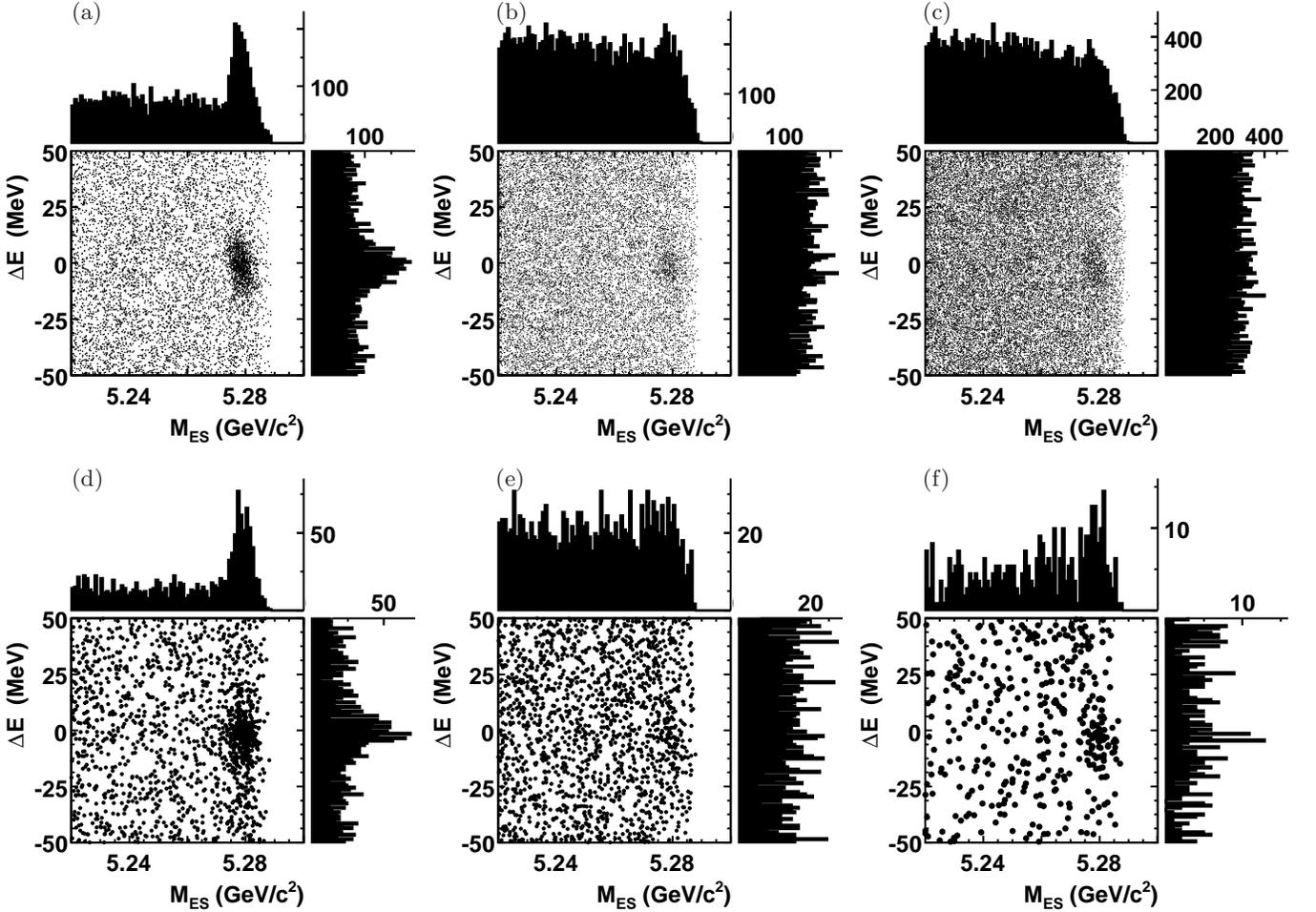} 
\caption{ 
    Scatter plots of $\MES$-$\DELTAE$ for the six newly observed
    $B$-meson decay modes:
    (a) ${B^-}{\rightarrow}{D^0}{p\pbar}{\pi^-}$,  
    (b) ${\Bbar^0}{\rightarrow}{D^0}{p\pbar}{\pi^-}{\pi^+}$,  
    (c) ${B^-}{\rightarrow}{D^+}{p\pbar}{\pi^-}{\pi^-}$, 
    (d) ${B^-}{\rightarrow}{D^{\ast0}}{p\pbar}{\pi^-}$,  
    (e) ${\Bbar^0}{\rightarrow}{D^{\ast0}}{p\pbar}{\pi^-}{\pi^+}$, 
    and (f) ${B^-}{\rightarrow}{D^{\ast+}}{p\pbar}{\pi^-}{\pi^-}$. 
    The first row of plots is related to the second by the exchange of
    the charm meson $D{\leftrightarrow}D^\ast$.  The decay chain
    involving ${D^0}{\rightarrow}{K^-}{\pi^+}$ or
    ${D^+}{\rightarrow}{K^-}{\pi^+}{\pi^+}$ is shown.  For (d, e, f),
    the decay chain involves ${D^{\ast0}}{\rightarrow}{D^0}{\pi^0}$ or
    ${D^{\ast+}}{\rightarrow}{D^0}{\pi^+}$.  The $\MES$ projection, in
    $1\mevcc$ bins, is given above the scatter plot; the $\DELTAE$, in
    $1\mev$ bins, on the right. For the projection plots, no selection
    is made on the complementary variable.
    }\label{fig:scatter} 
\end{figure*} 

\noindent
The $B$ meson beam-energy-substituted mass, $\MES$, and the difference
between its energy and the beam energy, $\DELTAE$, are defined with
the quantities in the lab frame:
\begin{equation} 
\begin{array} {ccl}
    \MES &=& \displaystyle{ 
        \sqrt{
        \frac{(s + 4\,\boldsymbol{P}_{\!B}{\cdot}\boldsymbol{P}_0)^2}
             {4\,(E_0)^2} - (\boldsymbol{P}_{\!B})^2  
            } 
        } \\ 
    \DELTAE &=& \displaystyle{
        \frac{Q_{\!\,B} \cdot Q_0}{\sqrt{s}} - 
        \frac{\sqrt{s}}{2} 
        }.
\end{array} 
\end{equation} 

\noindent
The four-momentum vectors $Q_{\!\,B}{=}(E_B,\boldsymbol{P}_{\!B})$ and
$Q_0{=}(E_0,\boldsymbol{P}_0)$ represent the $B$-meson candidate and
the ${e^+e^-}$ system, respectively.   The two variables, when
expressed in terms of center-of-mass quantities (denoted by
asterisks), take the more familiar form,
$\MES{=}\sqrt{s/4-(\boldsymbol{P}^{\,\ast}_{\!B})^2}$ and
$\DELTAE{=}E^{\,\ast}_{B}{-}\sqrt{s}/2$.

The $\MES$-$\DELTAE$ distributions for the events passing the final
selection are given for the six newly observed modes in
Fig.~\ref{fig:scatter}. Each point represents a candidate in an event.
For many of the modes, a dense concentration of events is visible near
$\MES{=}5.28\gevcc$, the PDG $B$-meson mass \cite{Amsler:2008zzb}, and
$\DELTAE{=}0$, as expected for signal events.  The uniform
distribution of events over the entire plane away from the signal area
is indicative of the general smoothness of the background event
distribution.

The $\MES$-$\DELTAE$ plots are given in a box region of
$5.22{<}\MES{<}5.30\gevcc$ and $|\DELTAE|{<}50\mev$.  This box is
large enough to provide a sufficient sideband region for each variable
where no signal events reside.  It is also small enough to exclude
possible contamination from other similarly related $B$-meson decay
modes.

For the purpose of plotting $\MES$ and $\DELTAE$ individually, the box
region is divided into a signal and a sideband region.  The $\MES$
signal region is within $2.5\,\sigma_{\MES}$ of the mean value of the
Gaussian function describing it and likewise for $\DELTAE$.
Similarly, the $\MES$ sideband region is outside $4\,\sigma_{\MES}$ of
the mean value and likewise for $\DELTAE$.  The resolutions range from
$2.2$ to $2.5\mevcc$ for $\sigma_{\MES}$ and $8$ to $10\mev$ for
$\sigma_{\DELTAE}$.  The signal box is the intersection of the $\MES$
and the $\DELTAE$ signal regions.

\subsection{Fit procedure} 
\label{sec:fit} 

\noindent
The signal yield is obtained by fitting the joint $\MES$-$\DELTAE$
distribution using a fit function in the
framework of the extended maximum likelihood technique
\cite{Barlow:1990vc}.  The likelihood value for $N$ observed events,
\begin{equation} 
    L(\hat{N},\hat{\Omega}) = \frac{e^{-\hat{N}}}{N!} 
    \,\prod^{N}_{i=1} P(y_i;\hat{N},\hat{\Omega}),
\end{equation} 

\noindent
is a function of the yield estimate $\hat{N}$ and the set of
parameters $\hat{\Omega}$.  The $y_i$ is the pair of $\MES$ and
$\DELTAE$ values for the $B$-meson candidate in the $i^\textrm{th}$
event and $P$ is described below.  The quantity $L$ is maximized
\cite{James:1975dr, Verkerke, Brun} with respect to its arguments.

The fit function is the sum of two terms 
\begin{equation} 
    P(y_i;\hat{N},\hat{\Omega}){=} 
    N_\textrm{sig}P_\textrm{sig}(y_i;{\Omega_\textrm{sig}}){+} 
    N_\textrm{bgd}P_\textrm{bgd}(y_i;{\Omega_\textrm{bgd}}),
\label{eq:PDF} 
\end{equation} 

\noindent
which correspond to the signal and the background component,
respectively.  For each component function, $P_\alpha$ is the
two-dimensional function, $N_\alpha$ the yield, and $\Omega_\alpha$ the
parameters.  The arguments of the function components are related to the
quantities in Eq.~(\ref{eq:PDF}) by $\hat{N}{=}\sum_{\beta}N_{\beta}$
and $\hat{\Omega}{=}\bigcup_\beta{\Omega}_\beta$.

Each function component $P_\alpha$ is written as the product of 
functions in $\MES$ and $\DELTAE$ since the variables are largely
uncorrelated.  (The signal bias due to the small correlation is
treated as a systematic uncertainty.)  The distributions for signal
events peak in each variable, so $P_\textrm{sig}$ is the product of
functions composed of a Gaussian core and a power-law tail
\cite{CBshape}.  The background event distribution varies smoothly, so
$P_\textrm{bgd}$ is the product of a threshold function
\cite{Aubert:2001xs} for $\MES$ that vanishes at approximately
$5.29\gevcc$ and a second-order Chebyshev polynomial for $\DELTAE$.

The following function parameters are fixed to the values found by
fitting the signal MC distributions: the $\DELTAE$ Gaussian width for
$P_\textrm{sig}$, the $\MES$ Gaussian width for $P_\textrm{sig}$, the
$\MES$ power-law tail parameters for $P_\textrm{sig}$, and the $\MES$
end-point parameter for $P_\textrm{bgd}$.  Two exceptions are given
after the detailed fit example.

\begin{figure}[tbp!] 
\centering 
\includegraphics[width=0.33\textwidth]{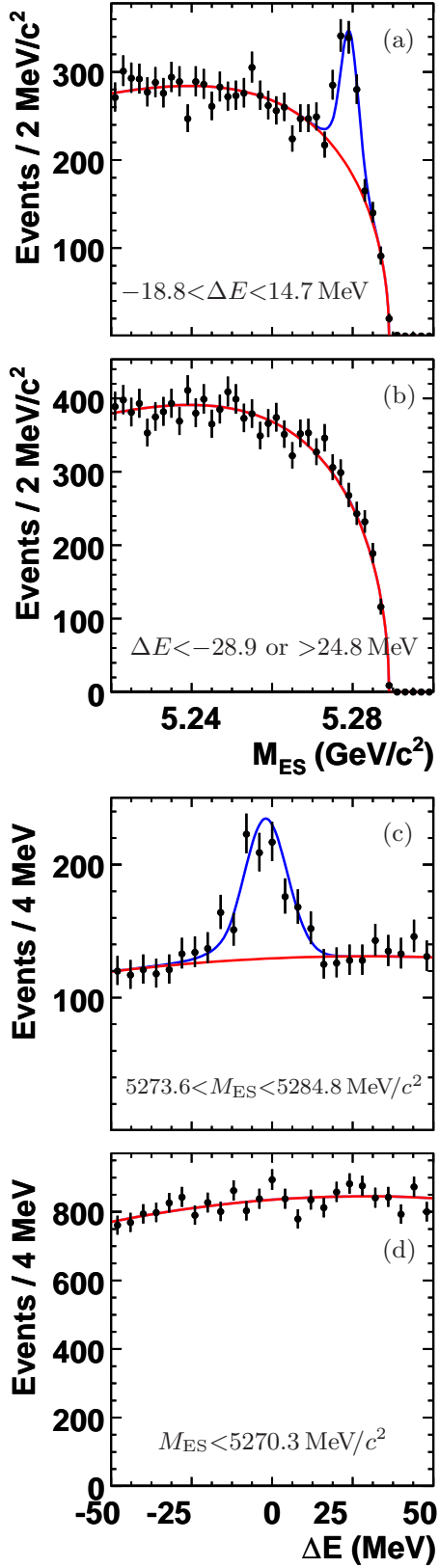} 
\caption{ 
    Fit details for ${B^-}{\rightarrow}{D^+}{p\pbar}{\pi^-}{\pi^-}$,
    ${D^+}{\rightarrow}{K^-}{\pi^+}{\pi^+}$: $\MES$ and $\DELTAE$
    distributions in regions as noted on the plots.  For (a, c) the
    top curve is the sum of $P_\textrm{sig}$ and $P_\textrm{bgd}$ and
    the bottom curve is the latter; for (b, d) the curve is
    $P_\textrm{bgd}$.   
    }\label{fig:examplefit}
\end{figure} 

A detailed example of the fit results is given in
Fig.~\ref{fig:examplefit} for the decay chain
${B^-}{\rightarrow}{D^+}{p\pbar}{\pi^-}{\pi^-}$,
${D^+}{\rightarrow}{K^-}{\pi^+}{\pi^+}$.  The plots in
Figs.~\ref{fig:examplefit}a and \ref{fig:examplefit}b show the $\MES$
distributions for the $\DELTAE$ signal and the $\DELTAE$ sideband
region, respectively.  Likewise, Figs.~\ref{fig:examplefit}c and
\ref{fig:examplefit}d show the respective $\DELTAE$ distributions for
the analogous $\MES$ regions.  The fit function projections describe
the distributions in the sideband regions well
(Figs.~\ref{fig:examplefit}b, \ref{fig:examplefit}d), which gives us
confidence that the background event distribution inside the signal
box are also modeled well. 

The first exception to the fit procedure described above applies to
the mode ${B^-}{\rightarrow}{D^{\ast0}}{p\pbar}{\pi^-}$.  A term is
added to Eq.~(\ref{eq:PDF}) to account for the sizable contamination
from the mode ${\Bbar^0}{\rightarrow}{D^{\ast+}}{p\pbar}{\pi^-}$.  The
fit function $P_\textrm{peak}$ is the same form as $P_\textrm{sig}$
with its parameters fixed to the values found by fitting the MC
sample.  The normalization $N_\textrm{peak}$ is based on the branching
fraction measured in this paper.

The second exception applies to four decay chains whose fits do not
converge: ${\Bbar^0}{\rightarrow}{D^0}{p\pbar}{\pi^-}{\pi^+}$,
$D^0{\rightarrow}K^-\pi^+\pi^0$;
${\Bbar^0}{\rightarrow}{D^0}{p\pbar}{\pi^-}{\pi^+}$,
$D^0{\rightarrow}K^-\pi^+\pi^-\pi^+$;
${\Bbar^0}{\rightarrow}{D^{\ast0}}{p\pbar}{\pi^-}{\pi^+}$,
$D^{\ast0}{\rightarrow}D^0\pi^0$, $D^0{\rightarrow}K^-\pi^+\pi^0$; and
${\Bbar^0}{\rightarrow}{D^{\ast0}}{p\pbar}{\pi^-}{\pi^+}$,
$D^{\ast0}{\rightarrow}D^0\pi^0$,
$D^0{\rightarrow}K^-\pi^+\pi^-\pi^+$.  Two changes are made: the
Gaussian parameters are fixed to the values found in the
${D^0}{\rightarrow}{K^-}{\pi^+}$ measurement, and the $\MES$ end-point
parameter is floated.  The fits converge after the changes. 

\section{BRANCHING FRACTIONS} 
\label{sec:bf} 

\noindent
This section presents the $B$-meson branching fractions $\mathcal{B}$.
Section~\ref{sec:fitresults} shows the fits to the $\MES$-$\DELTAE$
distributions.  Sections~\ref{sec:formula} and \ref{sec:ratios} gives
the $\mathcal{B}$ values and their ratios, respectively.  Throughout
this section, we simply state and use the systematic uncertainties of
Sec.~\ref{sec:syst}. 

\subsection{\boldmath Fits of $\protect\MES$-$\protect\DELTAE$ 
distributions} 
\label{sec:fitresults} 

\begin{figure*}[tbp!] 
\centering 
\includegraphics[width=1\textwidth]{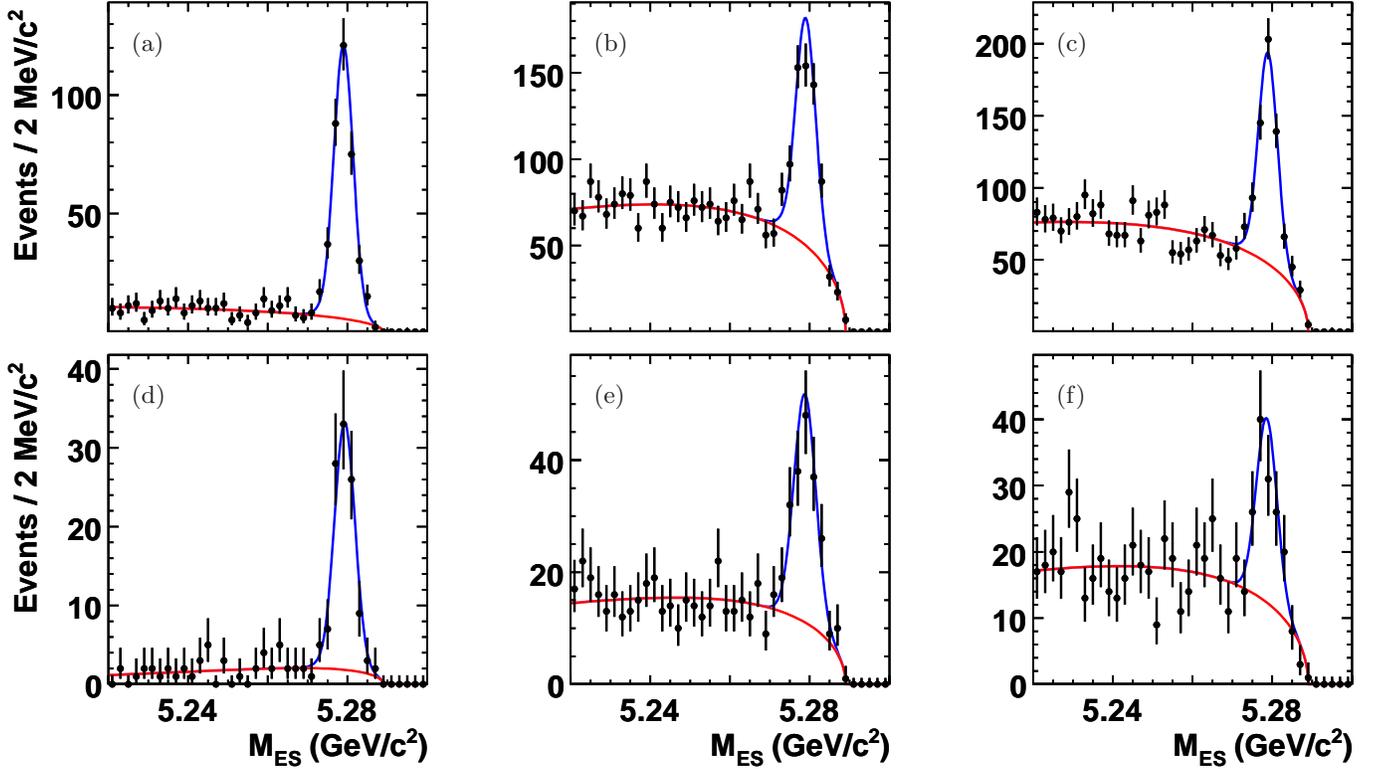} 
\caption{ 
    $\MES$ fit projections for the three-body modes: 
    (a--c) ${\Bbar^0}{\rightarrow}{D^0}{p\pbar}$ and  
    (d--f) ${\Bbar^0}{\rightarrow}{D^{\ast0}}{p\pbar}$,
    where
    (a, d) are reconstructed via ${D^0}{\rightarrow}{K^-}{\pi^+}$, 
    (b, e) ${D^0}{\rightarrow}{K^-}{\pi^+}{\pi^0}$, and
    (c, f) ${D^0}{\rightarrow}{K^-}{\pi^+}{\pi^-}{\pi^+}$; and
    (d--f) ${D^{\ast0}}{\rightarrow}{D^0}{\pi^0}$.
    Events with $\DELTAE$ within $2.5\sigma$ of the mean value of 
    the Gaussian function are shown.  The top curve is the sum
    of $P_\textrm{sig}$ and $P_\textrm{bgd}$ and the bottom curve 
    is the latter.   
    }\label{fig:mes3body} 
\end{figure*}

\begin{figure*}[tbp!] 
\centering 
\includegraphics[width=1\textwidth]{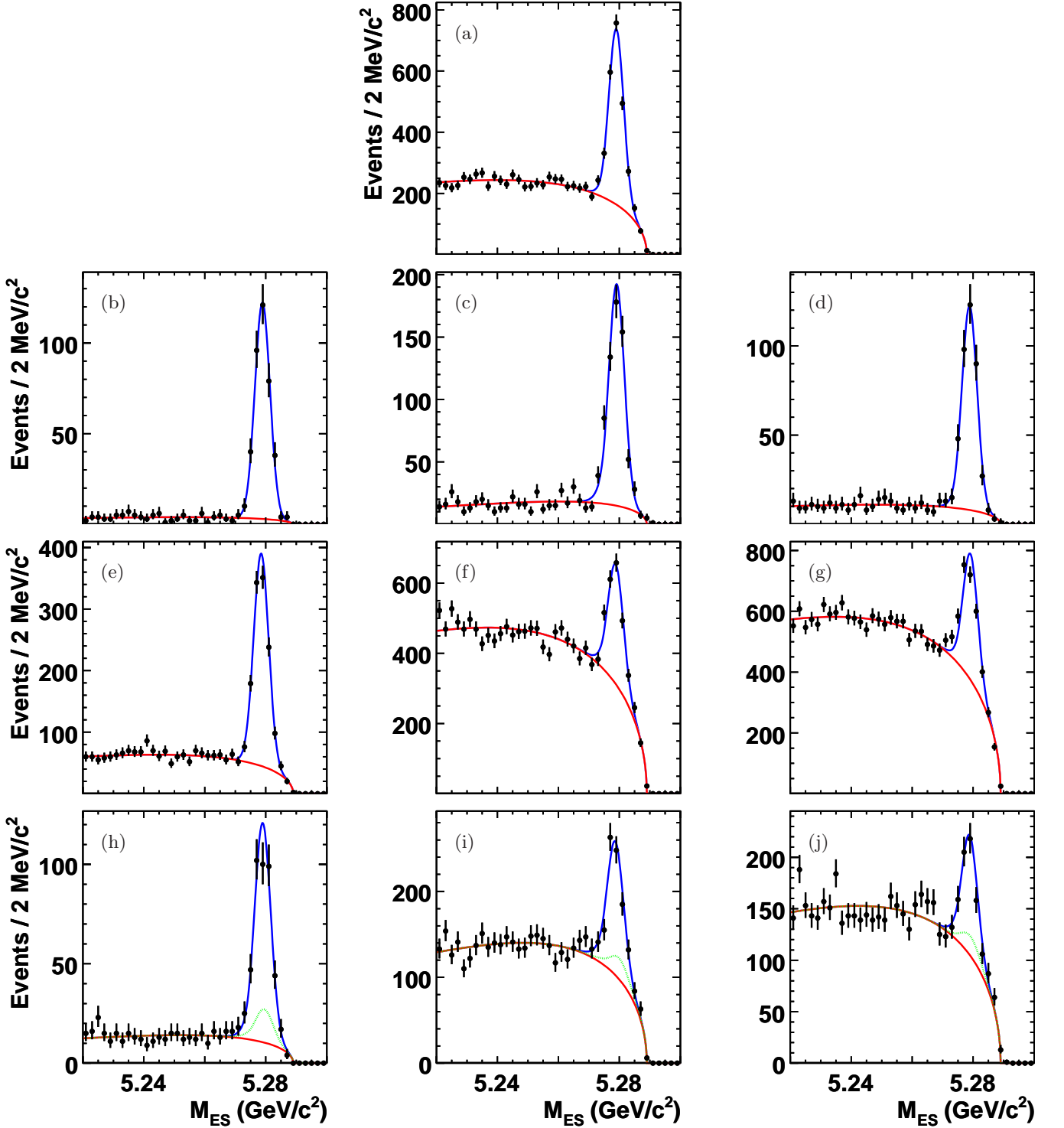} 
\caption{ 
    $\MES$ fit projections for the four-body modes: 
    (a)    ${\Bbar^0}{\rightarrow}{D^+}{p\pbar}{\pi^-}$,  
    (b--d) ${\Bbar^0}{\rightarrow}{D^{\ast+}}{p\pbar}{\pi^-}$, 
    (e--g) ${B^-}{\rightarrow}{D^0}{p\pbar}{\pi^-}$, and  
    (h--j) ${B^-}{\rightarrow}{D^{\ast0}}{p\pbar}{\pi^-}$, 
    where
    (a) is reconstructed via ${D^+}{\rightarrow}{K^-}{\pi^+}{\pi^+}$,
    (b, e, h) ${D^0}{\rightarrow}{K^-}{\pi^+}$, 
    (c, f, i) ${D^0}{\rightarrow}{K^-}{\pi^+}{\pi^0}$, and
    (d, g, j) ${D^0}{\rightarrow}{K^-}{\pi^+}{\pi^-}{\pi^+}$; and
    (b--d) ${D^{\ast+}}{\rightarrow}{D^0}{\pi^+}$ and
    (h--j) ${D^{\ast0}}{\rightarrow}{D^0}{\pi^0}$.
    Events with $\DELTAE$ within $2.5\sigma$ of the Gaussian mean 
    value are shown. 
    For (a--g) the top curve is the sum of $P_\textrm{sig}$
    and $P_\textrm{bgd}$ and the bottom curve is the latter;
    for (h--j) the middle curve is the sum of
    $P_\textrm{peak}$ and $P_\textrm{bgd}$.   
    }\label{fig:mes4body} 
\end{figure*}

\begin{figure*}[tbp!] 
\centering 
\includegraphics[width=1\textwidth]{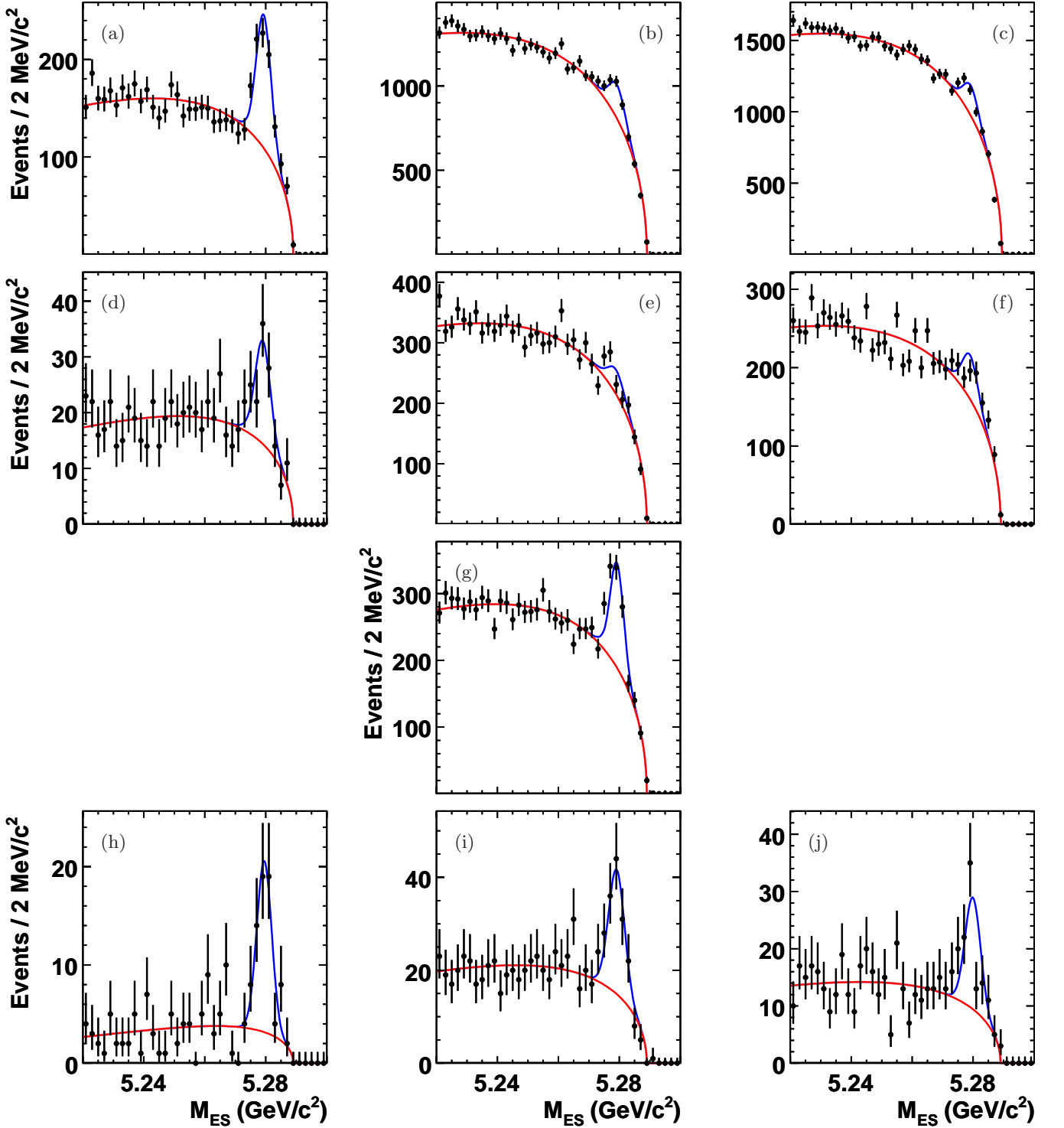} 
\caption{ 
    $\MES$ fit projections for the five-body modes: 
    (a--c) ${\Bbar^0}{\rightarrow}{D^0}{p\pbar}{\pi^-}{\pi^+}$, 
    (d--f) ${\Bbar^0}{\rightarrow}{D^{\ast0}}{p\pbar}{\pi^-}{\pi^+}$, 
    (g)    ${B^-}{\rightarrow}{D^+}{p\pbar}{\pi^-}{\pi^-}$, and 
    (h--j) ${B^-}{\rightarrow}{D^{\ast+}}{p\pbar}{\pi^-}{\pi^-}$,
    where
    (a, d, h) are reconstructed via ${D^0}{\rightarrow}{K^-}{\pi^+}$, 
    (b, e, i) ${D^0}{\rightarrow}{K^-}{\pi^+}{\pi^0}$,
    (c, f, j) ${D^0}{\rightarrow}{K^-}{\pi^+}{\pi^-}{\pi^+}$, and
    (g)       ${D^+}{\rightarrow}{K^-}{\pi^+}{\pi^+}$; and
    (b--d)    ${D^{\ast+}}{\rightarrow}{D^0}{\pi^+}$ and
    (h--j)    ${D^{\ast0}}{\rightarrow}{D^0}{\pi^0}$.   
    Events with $\DELTAE$ within $2.5\sigma$ of the Gaussian mean 
    value are shown. 
    The top curve is the sum of $P_\textrm{sig}$ and the
    $P_\textrm{bgd}$ and the bottom curve is the latter.  We
    note that the plots in (b, c, e, f) had difficulties with fit
    convergence; see text.   
    }\label{fig:mes5body} 
\end{figure*}

\noindent
The $\MES$ distributions for the events in the $\DELTAE$ signal region
for three-, four-, and five-body modes are given in
Figs.~\ref{fig:mes3body}--\ref{fig:mes5body}, respectively.  For all
$B$-meson decay modes, the 
decay chains involving ${D^0}{\rightarrow}{K^-}{\pi^+}$ or
${D^+}{\rightarrow}{K^-}{\pi^+}{\pi^+}$ show a peak.

The fit function projection in each plot describes the data well,
except for the four decay chains corresponding to
Figs.~\ref{fig:mes5body}b, \ref{fig:mes5body}c, \ref{fig:mes5body}d,
and \ref{fig:mes5body}f, which had difficulties with fit convergence
as noted in the previous section.  As we will see in
Sec.~\ref{sec:syst}, the yields from these decay chains do not
contribute significantly to the $B$-meson branching fraction, which is
dominated by the value from ${D^0}{\rightarrow}{K^-}{\pi^+}$, because
of their relatively large systematic uncertainties.

The signal yields, given in Table~\ref{tab:bf_chain}, range from $50$
to $3500$ events per mode.  

\subsection{Branching-fraction calculation} 
\label{sec:formula} 

\noindent
The $B$-meson branching fraction for each of the twenty-six decay
chains, given in Table~\ref{tab:bf_chain}, is given by 
\begin{equation} 
    \mathcal{B} = 
    \frac{1}{2\,N_{\!B\!\BbarSubscr}}
    \frac{1}{\mathcal{B}_{\Upsilon}\,\mathcal{B}_D\,\mathcal{B}_{D^{\ast}}}
    \frac{1}{\epsilon} 
    \Big(N_\textrm{sig} - N_\textrm{peak}\Big),
\label{eq:formula} 
\end{equation} 

\noindent
whose ingredients are as follows: the number of $B\Bbar$ pairs,
$N_{\!B\!\BbarSubscr}{=}455{\times}10^6$, the assumed
$\Upsilon{(4S)}{\rightarrow}B\Bbar^0$ or ${\rightarrow}B^+B^-$
branching fraction, $\mathcal{B}_{\Upsilon}{=}1/2$; the $D$-meson
branching fraction, $\mathcal{B}_D$ \cite{Amsler:2008zzb}; the
${D^\ast}$-meson branching fraction, $\mathcal{B}_{D^\ast}$
\cite{Amsler:2008zzb}; the reconstruction efficiency, $\epsilon$; the
signal yield, $N_\textrm{sig}$; and the measured contamination,
$N_\textrm{peak}$, using the $M(D)$-sideband data sample.  The
$\mathcal{B}_{D^\ast}$ is included only when a $D^\ast$ decay is
present in the decay chain.  The efficiency $\epsilon$ is determined
using the signal MC sample and decreases with the particle
multiplicity.  The mode ${\Bbar^0}{\rightarrow}{D^0}{p\pbar}$,
${D^0}{\rightarrow}{K^-}{\pi^+}$ has the highest value of $\epsilon$
at $19\%$ and
${\Bbar^0}{\rightarrow}{D^{\ast0}}{p\pbar}{\pi^-}{\pi^+}$,
${D^{\ast0}}{\rightarrow}{D^0}{\pi^0}$ and
${D^0}{\rightarrow}{K^-}{\pi^+}{\pi^-}{\pi^+}$ has the lowest at
$1\%$.

The $\mathcal{B}$ values, given in Table~\ref{tab:bf_mode}, are the
combinations \cite{Lyons:1988rp} of the above measurements using the
statistical and the systematic uncertainties.  All $\mathcal{B}$
values are significant with respect to their uncertainties.  For the
previously observed modes, the results are consistent with earlier
measurements.

\begin{table}[tbp!] 
\centering 
\caption{ 
    Intermediate values for Table~\ref{tab:bf_mode}: $B$-meson
    branching fractions for the decay chains.  $N_\textrm{sig}$ is
    the yield, $N_\textrm{peak}$ is the measured contamination
    (item xvii in Table~\ref{tab:syst_individual}), and
    $\epsilon$ is the reconstruction efficiency.  The
    uncertainties are statistical.  The rows marked by a
    dagger $\dag$ have large systematic uncertainties; see text.
    The charges of the pions are implied as well as the
    $D^{\ast0}{\rightarrow}D^0\pi^0$ and
    $D^{\ast+}{\rightarrow}D^0\pi^+$ decays, when applicable.
    }\label{tab:bf_chain} 
{\small 
\begin{tabular}{lr@{$\pm$}lrrcr@{$\pm$}l} 
\dbline 
    \multicolumn{1}{l}{$B$ modes, $D$ modes}
    & $N_\textrm{sig}$ 
    & $\sigma_\textrm{sig}$ 
    & $N_\textrm{peak}$
    & \multicolumn{1}{c}{$\epsilon$} 
    & 
    & $\mathcal{B}$
    & $\sigma_\textrm{stat}$\Bang
    \\ 
    & \multicolumn{2}{c}{} 
    & \multicolumn{1}{c}{} 
    & \multicolumn{1}{c}{(\%)} 
    & 
    & \multicolumn{2}{c}{($10^{-4}$)} 
    \\ 
\sgline 
    $\Bbar^0{\rightarrow}{D^0}{p\pbar}$, $K\pi$                            &$351$  &$20$ &$7.6$  &$19.0$ &&$1.02$ &$0.06$\\ 
    $\Bbar^0{\rightarrow}{D^0}{p\pbar}$, $K\pi\pi^0$                       &$431$  &$28$ &$24$   &$7.0$  &&$0.95$ &$0.06$\\ 
    $\Bbar^0{\rightarrow}{D^0}{p\pbar}$, $K\pi\pi\pi$                      &$448$  &$27$ &$10$   &$9.9$  &&$1.21$ &$0.07$\\ 
    $\Bbar^0{\rightarrow}{D^{\ast0}}{p\pbar}$, $K\pi$                      &$110$  &$12$ &$-1.4$ &$9.4$  &&$1.08$ &$0.12$\\ 
    $\Bbar^0{\rightarrow}{D^{\ast0}}{p\pbar}$, $K\pi\pi^0$                 &$148$  &$15$ &$3.9$  &$3.2$  &&$1.17$ &$0.12$\\ 
    $\Bbar^0{\rightarrow}{D^{\ast0}}{p\pbar}$, $K\pi\pi\pi$                &$95$   &$14$ &$5.5$  &$5.2$  &&$0.76$ &$0.12$\\ 
    $\Bbar^0{\rightarrow}{D^+}{p\pbar}{\pi^-}$, $K\pi\pi$                  &$1816$ &$53$ &$55$   &$12.6$ &&$3.32$ &$0.10$\\ 
    $\Bbar^0{\rightarrow}{D^{\ast+}}{p\pbar}{\pi^-}$, $K\pi$               &$392$  &$21$ &$2.3$  &$6.8$  &&$4.79$ &$0.26$\\ 
    $\Bbar^0{\rightarrow}{D^{\ast+}}{p\pbar}{\pi^-}$, $K\pi\pi^0$          &$601$  &$28$ &$21$   &$3.1$  &&$4.53$ &$0.22$\\ 
    $\Bbar^0{\rightarrow}{D^{\ast+}}{p\pbar}{\pi^-}$, $K\pi\pi\pi$         &$378$  &$22$ &$20$   &$3.7$  &&$3.92$ &$0.24$\\ 
    $B^-{\rightarrow}{D^0}{p\pbar}{\pi^-}$, $K\pi$                         &$1078$ &$38$ &$13$   &$15.9$ &&$3.79$ &$0.14$\\ 
    $B^-{\rightarrow}{D^0}{p\pbar}{\pi^-}$, $K\pi\pi^0$                    &$1176$ &$54$ &$41$   &$5.5$  &&$3.34$ &$0.16$\\ 
    $B^-{\rightarrow}{D^0}{p\pbar}{\pi^-}$, $K\pi\pi\pi$                   &$1296$ &$57$ &$33$   &$7.8$  &&$4.38$ &$0.20$\\ 
    $B^-{\rightarrow}{D^{\ast0}}{p\pbar}{\pi^-}$, $K\pi$                   &$328$  &$22$ &$2.1$  &$7.7$  &&$3.86$ &$0.26$\\ 
    $B^-{\rightarrow}{D^{\ast0}}{p\pbar}{\pi^-}$, $K\pi\pi^0$              &$482$  &$35$ &$47$   &$2.9$  &&$3.99$ &$0.32$\\ 
    $B^-{\rightarrow}{D^{\ast0}}{p\pbar}{\pi^-}$, $K\pi\pi\pi$             &$343$  &$31$ &$32$   &$4.0$  &&$3.37$ &$0.34$\\ 
    $\Bbar^0{\rightarrow}{D^0}{p\pbar}{\pi^-}{\pi^+}$, $K\pi$              &$438$  &$32$ &$7.7$  &$8.2$  &&$2.97$ &$0.22$\\ 
    $\Bbar^0{\rightarrow}{D^0}{p\pbar}{\pi^-}{\pi^+}$, $K\pi\pi^0$         &$663$  &$65$ &$160$  &$2.9$  &&$2.83$ &$0.36^{\dag}$\Bang\\ 
    $\Bbar^0{\rightarrow}{D^0}{p\pbar}{\pi^-}{\pi^+}$, $K\pi\pi\pi$        &$770$  &$68$ &$40$   &$3.8$  &&$5.28$ &$0.48^{\dag}$\Bang\\ 
    $\Bbar^0{\rightarrow}{D^{\ast0}}{p\pbar}{\pi^-}{\pi^+}$, $K\pi$        &$61$   &$12$ &$1.8$  &$2.9$  &&$1.87$ &$0.38$\\ 
    $\Bbar^0{\rightarrow}{D^{\ast0}}{p\pbar}{\pi^-}{\pi^+}$, $K\pi\pi^0$   &$142$  &$32$ &$37$   &$1.3$  &&$2.19$ &$0.66^{\dag}$\Bang\\ 
    $\Bbar^0{\rightarrow}{D^{\ast0}}{p\pbar}{\pi^-}{\pi^+}$, $K\pi\pi\pi$  &$163$  &$30$ &$13$   &$1.3$  &&$4.93$ &$0.99^{\dag}$\Bang\\ 
    $B^-{\rightarrow}{D^+}{p\pbar}{\pi^-}{\pi^-}$, $K\pi\pi$               &$475$  &$37$ &$6.6$  &$6.7$  &&$1.66$ &$0.13$\\ 
    $B^-{\rightarrow}{D^{\ast+}}{p\pbar}{\pi^-}{\pi^-}$, $K\pi$            &$57$   &$\phantom{0}9$ &$-12$ &$2.9$ &&$1.98 $&$0.26$\\ 
    $B^-{\rightarrow}{D^{\ast+}}{p\pbar}{\pi^-}{\pi^-}$, $K\pi\pi^0$\bang  &$94$   &$14$ &$-0.6$ &$1.3$ &&$1.82$ &$0.27$\\ 
    $B^-{\rightarrow}{D^{\ast+}}{p\pbar}{\pi^-}{\pi^-}$, $K\pi\pi\pi$\BANG &$66$   &$12$ &$4.8$  &$1.5$ &&$1.61$ &$0.32$\\ 
\dbline 
\end{tabular} 
} 
\end{table} 

\begin{table*}[tbp!] 
\centering 
\caption{ 
    Main results of this paper: $B$-meson branching fractions for the
    ten modes.  Also given are the values of $\chi^2$, the degrees of
    freedom (DOF), and the $\chi^2$ probabilities for the averaging of
    the results from Table~\ref{tab:bf_chain}.  The measurements are
    consistent with the previous results.
    }\label{tab:bf_mode} 
{\small 
\begin{tabular}{
    clr@{$\pm$}l@{$\pm$}l  
    p{0.03\textwidth}cc  
    p{0.02\textwidth}r@{$\pm$}l@{$\pm$}l  
    p{0.01\textwidth}r@{$\pm$}l@{$\pm$}l
}
\dbline 
\multicolumn{1}{l}{$N$-body} 
    & $B$-meson decay mode
    & $\mathcal{B}$\phantom{,} 
    & $\sigma_\textrm{stat}$  
    & $\sigma_\textrm{syst}$  
    & 
    & $\chi^2/\textrm{DOF}$ &  $\textrm{Prob}(\chi^2)$ 
    & 
    & \multicolumn{3}{c}{$\mathcal{B}$ from Refs.\ \onlinecite{Anderson:2000tz,Abe:2002tw}} 
    & 
    & \multicolumn{3}{c}{$\mathcal{B}$ from Ref.\ \onlinecite{Aubert:2006qx}} 
    \\ 
    & 
    & \multicolumn{3}{c}{($10^{-4}$)} 
    & \multicolumn{2}{c}{\phantom{($10^{-4}$)}} 
    & (\%) 
    & 
    & \multicolumn{3}{c}{($10^{-4}$)} 
    & 
    & \multicolumn{3}{c}{($10^{-4}$)} 
    \\ 
\sgline 
    \multicolumn{1}{l}{Three-body}&
        $\Bbar^0\rightarrow{D^0}p\pbar$
            &$1.02$ &$0.04$ &$0.06$ &&$4.3/2$ &$12$ 
            &&$1.18$ &$0.15$ &$0.16$ \cite{Abe:2002tw} 
            &&$1.13$ &$0.06$ &$0.08$ \\
        $''$&
        $\Bbar^0\rightarrow{D^{\ast0}}p\pbar$
            &$0.97$ &$0.07$ &$0.09$ &&$4.1/2$ &$13$ 
            &&$1.20$ &$^{0.33}_{0.29}$ &$0.21$ \cite{Abe:2002tw} 
            &&$1.01$ &$0.10$ &$0.09$ \\
    \multicolumn{1}{l}{Four-body}&
        $\Bbar^0\rightarrow{D^+}p\pbar\pi^-$
            &$3.32$ &$0.10$ &$0.29$ &&- &-    
            &&\multicolumn{3}{c}{-} 
            &&$3.38$ &$0.14$ &$0.29$ \\
        $''$& 
        $\Bbar^0\rightarrow{D^{\ast+}}p\pbar\pi^-$
            &$4.55$ &$0.16$ &$0.39$ &&$1.2/2$ &$54$ 
            &&$6.5\phantom{0}$  &$^{1.3}_{1.2}\phantom{0}$
                &$1.0\phantom{0}$ \cite{Anderson:2000tz} 
            &&$4.81$ &$0.22$ &$0.44$ \\
        $''$&
        $B^-\rightarrow{D^0}p\pbar\pi^-$
            &$3.72$ &$0.11$ &$0.25$ &&$3.4/2$ &$19$ 
            &&\multicolumn{3}{c}{-}
            &&\multicolumn{3}{c}{-} \\
        $''$&
        $B^-\rightarrow{D^{\ast0}}p\pbar\pi^-$
            &$3.73$ &$0.17$ &$0.27$ &&$0.5/2$ &$79$ 
            &&\multicolumn{3}{c}{-}
            &&\multicolumn{3}{c}{-} \\
    \multicolumn{1}{l}{Five-body}&
        $\Bbar^0\rightarrow{D^0}p\pbar\pi^-\pi^+$
            &$2.99$ &$0.21$ &$0.45$ &&$0.3/2$ &$85$ 
            &&\multicolumn{3}{c}{-}
            &&\multicolumn{3}{c}{-} \\
        $''$&
        $\Bbar^0\rightarrow{D^{\ast0}}p\pbar\pi^-\pi^+$
            &$1.91$ &$0.36$ &$0.29$ &&$0.5/2$ &$78$ 
            &&\multicolumn{3}{c}{-}
            &&\multicolumn{3}{c}{-} \\
        $''$&
        $B^-\rightarrow{D^+}p\pbar\pi^-\pi^-$
            &$1.66$ &$0.13$ &$0.27$ &&- &-    
            &&\multicolumn{3}{c}{-}
            &&\multicolumn{3}{c}{-} \\
        $''$&
        $B^-\rightarrow{D^{\ast+}}p\pbar\pi^-\pi^-$
            &$1.86$ &$0.16$ &$0.19$ &&$0.2/2$ &$91$ 
            &&\multicolumn{3}{c}{-}
            &&\multicolumn{3}{c}{-} \\
\dbline 
\end{tabular} 
} 
\end{table*} 

\subsection{Branching-fraction ratios} 
\label{sec:ratios} 

\noindent
Table~\ref{tab:bf_ratio} gives the ratio of the branching fractions
$\mathcal{B}$ for modes related by $D{\leftrightarrow}D^\ast$,
$D^{(\!\ast\!)0}{\leftrightarrow}D^{(\!\ast\!)+}$, and the addition of
$\pi$.  These ratios show four patterns:

(i) The ratios are roughly unity for the modes related by the
spin of the charm mesons, $D{\leftrightarrow}{D^\ast}$.  This result
suggests that the additional degrees of freedom due to the ${D^\ast}$
polarization vector do not significantly modify the production rate.

(ii) The ratio is roughly unity for the modes related by the
charge of the charm mesons,
${D^{(\!\ast\!)+}}{\leftrightarrow}{D^{(\!\ast\!)0}}$,

(iii) The ratio for the four-body mode to that of the corresponding
three-body mode with one fewer pion is about four.

(iv) The ratio for the five-body mode to that of the corresponding
four-body mode with one fewer pion is about one-half.

The patterns (iii, iv) imply
$\mathcal{B}_\textrm{\,3-body}{<}\mathcal{B}_\textrm{\,5-body}{<}\mathcal{B}_\textrm{\,4-body}$. 

\begin{table}[tbp!] 
\centering 
\caption{ 
    Ratios of $B$-meson branching fractions of the modes related by
    $D{\leftrightarrow}D^\ast$,
    $D^{(\!\ast\!)0}{\leftrightarrow}D^{(\!\ast\!)+}$, and 
    the addition of $\pi$.  The uncertainties are statistical. 
    }\label{tab:bf_ratio} 
{ 
\small 
\begin{tabular}{p{0.015\textwidth}p{2.5in}r@{${\pm}$}c} 
\dbline 
    \multicolumn{2}{l}{Ratio of the modes} & $R$\ \ & $\sigma_{R}$ \\ 
\sgline 
    \multicolumn{2}{l}{\textrm{Related by spin of charm meson}} \\
    &$\mathcal{B}({\Bbar^0}{\rightarrow}{D^{\ast0}}{p\pbar})$/$\mathcal{B}({\Bbar^0}{\rightarrow}{D^0}{p\pbar})$                               &$0.95$ &$0.08$ \\ 
    &$\mathcal{B}({\Bbar^0}{\rightarrow}{D^{\ast+}}{p\pbar}{\pi^-})$/$\mathcal{B}({\Bbar^0}{\rightarrow}{D^+}{p\pbar}{\pi^-})$                 &$1.37$ &$0.06$ \\ 
    &$\mathcal{B}({B^-}{\rightarrow}{D^{\ast0}}{p\pbar}{\pi^-})$/$\mathcal{B}({B^-}{\rightarrow}{D^0}{p\pbar}{\pi^-})$                         &$1.00$ &$0.05$ \\ 
    &$\mathcal{B}({\Bbar^0}{\rightarrow}{D^{\ast0}}{p\pbar}{\pi^-}{\pi^+})$/$\mathcal{B}({\Bbar^0}{\rightarrow}{D^0}{p\pbar}{\pi^-}{\pi^+})$   &$0.64$ &$0.13$ \\ 
    &$\mathcal{B}({B^-}{\rightarrow}{D^{\ast+}}{p\pbar}{\pi^-}{\pi^-})$/$\mathcal{B}({B^-}{\rightarrow}{D^+}{p\pbar}{\pi^-}{\pi^-})$           &$1.12$ &$0.13$ \\ 
    \multicolumn{2}{l}{\textrm{Related by charge of charm meson}}\\ 
    &$\mathcal{B}({B^-}{\rightarrow}{D^0}{p\pbar}{\pi^-})$/$\mathcal{B}({\Bbar^0}{\rightarrow}{D^+}{p\pbar}{\pi^-})$                           &$1.12$ &$0.05$ \\ 
    &$\mathcal{B}({B^-}{\rightarrow}{D^{\ast0}}{p\pbar}{\pi^-})$/$\mathcal{B}({\Bbar^0}{\rightarrow}{D^{\ast+}}{p\pbar}{\pi^-})$               &$0.82$ &$0.05$ \\ 
    &$\mathcal{B}({\Bbar^0}{\rightarrow}{D^0}{p\pbar}{\pi^-}{\pi^+})$/$\mathcal{B}({B^-}{\rightarrow}{D^+}{p\pbar}{\pi^-}{\pi^-})$             &$1.80$ &$0.19$ \\ 
    &$\mathcal{B}({\Bbar^0}{\rightarrow}{D^{\ast0}}{p\pbar}{\pi^-}{\pi^+})$/$\mathcal{B}({B^-}{\rightarrow}{D^{\ast+}}{p\pbar}{\pi^-}{\pi^-})$ &$1.03$ &$0.21$ \\ 
    \multicolumn{3}{l}{\textrm{Related by addition of pion to three-body modes}}\\ 
    &$\mathcal{B}({B^-}{\rightarrow}{D^{\ast0}}{p\pbar}{\pi^-})$/$\mathcal{B}({\Bbar^0}{\rightarrow}{D^{\ast0}}{p\pbar})$                      &$3.84$ &$0.33$ \\ 
    &$\mathcal{B}({B^-}{\rightarrow}{D^0}{p\pbar}{\pi^-})$/$\mathcal{B}({\Bbar^0}{\rightarrow}{D^0}{p\pbar})$                                  &$3.64$ &$0.18$ \\ 
    \multicolumn{3}{l}{\textrm{Related by addition of pion to four-body modes}}\\ 
    &$\mathcal{B}({B^-}{\rightarrow}{D^+}{p\pbar}{\pi^-}{\pi^-})$/$\mathcal{B}({\Bbar^0}{\rightarrow}{D^+}{p\pbar}{\pi^-})$                    &$0.50$ &$0.04$ \\ 
    &$\mathcal{B}({B^-}{\rightarrow}{D^{\ast+}}{p\pbar}{\pi^-}{\pi^-})$/$\mathcal{B}({\Bbar^0}{\rightarrow}{D^{\ast+}}{p\pbar}{\pi^-})$        &$0.41$ &$0.04$ \\ 
    &$\mathcal{B}({\Bbar^0}{\rightarrow}{D^0}{p\pbar}{\pi^-}{\pi^+})$/$\mathcal{B}({\Bbar^0}{\rightarrow}{D^0}{p\pbar}{\pi^-})$                &$0.80$ &$0.06$ \\ 
    &$\mathcal{B}({\Bbar^0}{\rightarrow}{D^{\ast0}}{p\pbar}{\pi^-}{\pi^+})$/$\mathcal{B}({\Bbar^0}{\rightarrow}{D^{\ast0}}{p\pbar}{\pi^-})$    &$0.51$ &$0.10$ \\ 
\dbline 
\end{tabular} 
} 
\end{table}  

\section{SYSTEMATIC UNCERTAINTIES} 
\label{sec:syst} 

\noindent
This section describes the systematic uncertainties for the $B$-meson
branching fraction measurement.  Section~\ref{sec:syst_ind} lists the
sources, and Sec.~\ref{sec:errormatrix} gives the error matrices.

\subsection{Sources} 
\label{sec:syst_ind} 

\noindent
The sources of systematic uncertainties, which are listed in
Table~\ref{tab:syst_individual}, can be organized as follows:

\begin{table}[tbp!] 
\centering 
\caption{ 
    Systematic uncertainty list for $B$-meson branching
    fractions. The ``$D$ modes''
    represents ${D^0}{\rightarrow}{K^-}{\pi^+}$,
    ${K^-}{\pi^+}{\pi^0}$, and ${K^-}{\pi^+}{\pi^-}{\pi^+}$; and
    ${D^+}{\rightarrow}{K^-}{\pi^+}{\pi^+}$.
    }\label{tab:syst_individual} 
{\small 
\begin{tabular}{lp{0.287\textwidth}l} 
\dbline 
    Item & Description & Uncertainty ($\%$)\\ 
\sgline 
    i    &Number of $B\Bbar$ pairs &$1.1$ \\ 
    ii   &$\mathcal{B}(\Upsilon{(4S)}$: for $\Upsilon{(4S)}{\rightarrow}B\Bbar$ &$3.2$ \\ 
    iii  &$\mathcal{B}({D})$: for $D$ modes &$1.8$, $4.4$, $3.2$, $3.6$ \\ 
    iv   &$\mathcal{B}({D^{\ast}})$: for ${D^{\ast}}{\rightarrow}{D^0}{\pi^0}$,~${D^{\ast+}}{\rightarrow}{D^0}{\pi^+}$ &$4.7$, $0.7$ \\ 
    v    &Charged particle reconstruction &$0.5$ \\ 
    vi   &${\pi^+}$ from ${D^{\ast+}}{\rightarrow}{D^0}{\pi^+}$ &$3.1$ \\ 
    vii  &${\pi^0}$ reconstruction &$3.0$ \\ 
    viii &Signal mode decay dynamics &$0.8$--$9.7$ \\ 
    ix   &Kaon and proton id using data &$1.5$--$2.5$ \\ 
    x    &Kaon id in $B\Bbar$ event topology &$0.5$ \\ 
    xi   &Proton id in $B\Bbar$ event topology &$1.0$ \\ 
    xii  &Fit function params: for $D$ modes &$1.3$, $2.8$, $5.7$, $3.4$ \\ 
    xiii &Signal fit function &$0.6$ \\ 
    xiv  &Backgrnd.\ fit function: for $D$ modes &$0.8$, $4.5$, $1.3$, $2.0$ \\ 
    xv   &$\MES$-$\DELTAE$ correlation &$0.4$--$2.2$ \\ 
    xvi  &Backgrnd.\ peaking in $\MES$ or $\DELTAE$ for all modes (marked $\dag$ in Table~\ref{tab:bf_chain}) & $0$--$5.5$ ($77$--$85$) \\ 
    xvii &Backgrnd. from baryonic modes &$0.5$--$13.5$ \\ 
\dbline 
\end{tabular} 
} 
\end{table} 

\noindent 
\begin{enumerate}\itemsep0pt 
    \item[$\bullet$](i) Counting of the number of $B\Bbar$ pairs, 
    \item[$\bullet$](ii--iv) Assumed branching fractions,
    \item[$\bullet$](v--xi) Reconstruction efficiencies,
    \item[$\bullet$](xii--xv) Fit functions and its parameters, and
    \item[$\bullet$](xvi--xvii) Backgrounds peaking in $\MES$ or $\DELTAE$.
\end{enumerate} 

\noindent
These contributions are described below.

(i) The number of $B\Bbar$ pairs used in the analysis is the
difference of the observed number of hadronic events and the expected
contribution from continuum events.  The latter is estimated using a
separate data sample taken $40\mev$ below the $\Upsilon(4S)$ peak.
The uncertainty of $1.1\%$ is mostly due to the difference in the
detection efficiencies for hadronic events in the data and the MC
samples. 

(ii) The $\Upsilon{(4S)}$ branching fraction is assumed to be equal
for ${B^0}{\Bbar^0}$ and ${B^+}{B^-}$.  The uncertainty of $3.2\%$ is
the difference of $1/2$ and the PDG value \cite{Amsler:2008zzb}.   

(iii, iv) The ${D}$- and ${D}^{\ast}$-meson branching fractions 
assume the PDG values \cite{Amsler:2008zzb}.  The
uncertainties of $1.3\%$, $3.7\%$, $2.5\%$, and $2.3\%$ are the
PDG uncertainties for ${D^0}{\rightarrow}{K^-}{\pi^+}$,
${K^-}{\pi^+}{\pi^0}$, ${K^-}{\pi^+}{\pi^-}{\pi^+}$, and
${D^+}{\rightarrow}{K^-}{\pi^+}{\pi^+}$, respectively; and $4.7\%$ and
$0.7\%$ for ${D^{\ast0}}{\rightarrow}{D^0}{\pi^0}$ and
${D^{\ast+}}{\rightarrow}{D^0}{\pi^+}$, respectively.

(v) The charged track reconstruction efficiency is evaluated using
${e^+e^-}{\rightarrow}\tau^+\tau^-$ events, where one tau decays
leptonically and the other hadronically.  The uncertainty of $0.5\%$
is due to the difference between the detection efficiency in the data
and the MC samples.

(vi) The reconstruction efficiency of low-energy charged pion from
${D^{\ast+}}{\rightarrow}{D^0}{\pi^+}$ decays is sufficiently
difficult, in comparison to other tracks, that item (v) cannot account
for its uncertainty.  Such a pion is often found using only the
silicon vertex tracker because its momentum is relatively low.  The
momentum dependence of pion identification is evaluated using the
helicity angle $\theta_\textrm{hel}$ distribution---the angle between
the pion direction in the ${D^{\ast+}}$ rest frame and the
${D^{\ast+}}$ boost direction---because the two quantities are highly
correlated.  Since the pions are produced symmetrically in
$\cos\theta_\textrm{hel}$, the observed asymmetry in the distribution
is indicative of the momentum dependence of the efficiency.  The
uncertainty of $3.1\%$ is due to the difference in the momentum
dependence in the data and the MC samples.

(vii) The $\pi^0$ reconstruction efficiency is evaluated using
$\tau^+\tau^-$ events as in item (v) with an uncertainty of $3.0\%$.

(viii) The signal $B$-candidate reconstruction efficiency is evaluated
using the MC samples.  Since these samples use the uniform phase-space
decay model while the reported baryonic decay dynamics
\cite[this~paper]{%
Lee:2004mg,%
Wang:2005fc,%
Medvedeva:2007zz,%
Wei:2007fg,%
Chen:2008jy,%
Aubert:2005gw,%
Anderson:2000tz,%
Abe:2002tw,%
Aubert:2006qx}
are far from uniform, corrections are made in the variables where the
strongest variation are seen---in bins of $M^2({p\pbar})$
vs.~$M^2({D^{(\!\ast\!)}}{p})$---using the data and the MC samples.
The uncertainties ranging from $0.8\%$ to $9.7\%$ are due to the
limited statistics of the samples.

(ix) The particle identification efficiencies for kaons and protons
are evaluated using the MC samples, which are then corrected using a
data sample rich in these hadrons.  The uncertainties ranging from
$1.5\%$ to $2.5\%$ are due to the sample statistics associated with the
correction procedure.  The sample, however, is dominated by the
continuum events whose event topology is different from $B\Bbar$
events.  Items (x, xi) account for the differences.

(x, xi) The kaon and proton identification efficiencies in the
$B\Bbar$ environment are evaluated using a data sample of
${D^{\ast+}}{\rightarrow}{D^0}{\pi^+}$,
${D^0}{\rightarrow}{K^-}{\pi^+}$ and $\Lambda{\rightarrow}p{\pi^-}$
decays, respectively.  The uncertainties of $0.5\%$ and $1.0\%$,
respectively, are due to the differences in the event topologies.

(xii) A subset of the fit function parameters is fixed when fitting
the $\MES$-$\DELTAE$ distributions in the data sample.  Such parameter
values are obtained by fitting the MC distributions, and they are
assigned an uncertainty from this fit.  The effect on the signal yield
is evaluated by fitting the data sample with the parameter value
shifted by $1\sigma$.  The procedure is repeated for each parameter in
the set.  The uncertainties of $1.3\%$, $2.8\%$, $5.7\%$, and $3.4$\%
for the modes with ${D^0}{\rightarrow}{K^-}{\pi^+}$,
${K^-}{\pi^+}{\pi^0}$, ${K^-}{\pi^+}{\pi^-}{\pi^+}$,
${K^-}{\pi^+}{\pi^+}$, and ${D^+}{\rightarrow}{K^-}{\pi^+}{\pi^+}$,
respectively, are the quadrature sum of the fractional yield changes.

(xiii) The choice of the signal fit function is evaluated using an
alternate function, a fourth-order polynomial.  The uncertainty of
$0.6\%$ is due to the yield difference with respect to the original
fit function.

(xiv) The choice of the background fit function is evaluated using a
more general fit function with the addition of another such component.
The uncertainties---$0.8\%$, $4.5\%$, $1.3\%$, and $2.0\%$ for the
modes with ${D^0}{\rightarrow}{K^-}{\pi^+}$, ${K^-}{\pi^+}{\pi^0}$,
${K^-}{\pi^+}{\pi^-}{\pi^+}$, and
${D^+}{\rightarrow}{K^-}{\pi^+}{\pi^+}$, respectively---are due to the
yield differences with respect to the original fit function.

(xv) The small correlation between the $\MES$ and $\DELTAE$
distributions introduces a bias in the signal yield.  This effect is
quantified by fitting pseudo-experiments. Each experiment contains a
background sample whose $\MES$ and $\DELTAE$ distributions are
produced according to $P_\textrm{bgd}$, and a signal MC sample from
the full detector simulation. The uncertainties ranging from $0.1\%$
to $1.8\%$ are from the deviation of $N_\mathrm{sig}$ to the mean of
the signal-yield distribution.

(xvi) Background events whose distributions peak either at
$\MES{=}5.28\gevcc$ or $\DELTAE{=}0$ can alter the signal yield.
For the ${B^-}{\rightarrow}D^{\ast0}{p\pbar}{\pi^-}$ measurement, the
variation of the normalization of the fit function for the
${\Bbar^0}{\rightarrow}D^{\ast+}{p\pbar}{\pi^-}$ contribution within
the experimental uncertainties has a negligible effect on the signal
yield.  For other $B$ decay modes, no such sources are found.
However, the $\MES$ distributions for a few cases feature a broad hump
with a width around $20\mevcc$ spanning nearly half of the signal box.
The effect of the presence of such a source is quantified by adding a
component $P_\textrm{peak}$ to the fit function whose parameters are
fixed except for the normalization.  Except for four decay
chains---those corresponding to Figs.~\ref{fig:mes5body}b,
\ref{fig:mes5body}c, \ref{fig:mes5body}e, and
\ref{fig:mes5body}f---uncertainties ranging from zero to $5.5\%$ are
obtained from the changes in yield when the additional component is
included.  For the mentioned exceptions, the uncertainties range from
$77\%$ to $85\%$.  As a consequence of the large uncertainties, these
four modes do not contribute significantly to the final results.

(xvii) Background events from baryonic modes without a $D$ meson are
evaluated using the data sample.  An example case where the final
states are identical is $B{\rightarrow}\Lambda_c\,\pbar{\pi^0}$,
$\Lambda_c{\rightarrow}{p}{K^-}{\pi^+}$ and
${\Bbar^0}{\rightarrow}{D^0}{p\pbar}$,
${D^0}{\rightarrow}{K^-}{\pi^+}{\pi^0}$.  For such a source, the
$M(D)$ distribution does not peak at the $D$ mass, so the
contamination can be quantified by repeating the analysis with the
$M(D)$-sideband region.  $N_\textrm{peak}$ is an additive correction
factor for $N_\textrm{sig}$ with uncertainties ranging from $0.5\%$ to
$13.5\%$ due to the sample statistics.

\subsection{Error matrices} 
\label{sec:errormatrix} 

\begin{table}[tbp!] 
\centering 
\caption{
    Systematic uncertainties (\%) combined for the $B$ modes.  For
    each $D$ mode, two columns are given.  The uncorrelated values are
    given on the left columns and the correlated on the right columns.
    The right columns exclude items (iii, iv) of
    Table~\ref{tab:syst_individual}. 
    }\label{tab:syst_combined} 
{\small 
\begin{tabular}{l rr rr rr rr} 
\dbline 
    \multirow{2}{*}{\backslashbox{$B$ mode}{${\ \ }D{\rightarrow}{\!\!\!\!}$}}
    &\multicolumn{2}{c}{\footnotesize${K^-}{\pi^+}$}  
    &\multicolumn{2}{c}{\footnotesize${K^-}{\pi^+}{\pi^0}$}  
    &\multicolumn{2}{c}{\footnotesize${K^-}{\pi^+}{\pi^-}{\pi^+}$\!\!\!\!}
    &\multicolumn{2}{c}{\footnotesize${K^-}{\pi^+}{\pi^+}$}  \\ 
    &\multicolumn{1}{r}{unc} &\multicolumn{1}{r}{\textrm{cor}} 
    &\multicolumn{1}{r}{unc} &\multicolumn{1}{r}{cor} 
    &\multicolumn{1}{r}{\phantom{00}unc} &\multicolumn{1}{r}{cor\phantom{00}} 
    &\multicolumn{1}{r}{unc} &\multicolumn{1}{r}{cor} 
    \\ 
\sgline 
    $\Bbar^0{\rightarrow}{D^0}p\pbar$                  &$2.7$  &$3.5$ &$5.5$  &$6.9$ &$4.7$  &$7.0\phantom{00}$ &- &-\\ 
    $\Bbar^0{\rightarrow}{D^{\ast0}}p\pbar$            &$2.2$  &$4.6$ &$8.6$  &$8.6$ &$8.7$  &$7.7\phantom{00}$ &- &-\\ 
    $\Bbar^0{\rightarrow}{D^+}p\pbar\pi^-$             &-      &-     &-      &-     &-      &   -\phantom{00}  &$5.7$ &$5.7$\\ 
    $\Bbar^0{\rightarrow}{D^{\ast+}}p\pbar\pi^-$       &$4.2$  &$6.3$ &$7.6$  &$8.6$ &$9.9$  &$8.9\phantom{00}$ &- &-\\ 
    $\Bbar^0{\rightarrow}{D^0}p\pbar\pi^-\pi^+$        &$14.5$ &$3.9$ &$81.2$ &$7.0$ &$77.3$ &$7.0\phantom{00}$ &- &-\\ 
    $\Bbar^0{\rightarrow}{D^{\ast0}}p\pbar\pi^-\pi^+$  &$13.8$ &$4.9$ &$86.3$ &$8.8$ &$85.4$ &$7.7\phantom{00}$ &- &-\\ 
    $B^-{\rightarrow}{D^0}p\pbar\pi^-$                 &$4.4$  &$4.2$ &$8.8$  &$7.2$ &$11.6$ &$7.3\phantom{00}$ &- &-\\ 
    $B^-{\rightarrow}{D^{\ast0}}p\pbar\pi^-$           &$6.6$  &$5.2$ &$8.4$  &$8.9$ &$10.5$ &$7.9\phantom{00}$ &- &-\\ 
    $B^-{\rightarrow}{D^+}p\pbar\pi^-\pi^-$            &-      &-     &-      &-     &-      &   -\phantom{00}  &$15.0$ &$5.9$ \\ 
    $B^-{\rightarrow}{D^{\ast+}}p\pbar\pi^-\pi^-$\Bang &$5.9$  &$6.8$ &$14.9$ &$9.0$ &$19.0$ &$9.2\phantom{00}$ &- &-\\ 
\dbline 
\end{tabular} 
} 
\end{table} 

\noindent
The error matrix, $\mathit{V}$, spanning the $D$ modes of a given $B$
mode is the sum of the statistical and systematic components
$\mathit{V}{=}\mathit{V}_\textrm{stat}{+}\mathit{V}_\textrm{syst}$.

The $\mathit{V}_\textrm{stat}$ is diagonal with elements
$(\sigma_{\textrm{stat},\alpha})^2$ (Table~\ref{tab:bf_chain}).

The $\mathit{V}_\textrm{syst}$ is the sum of a diagonal part and an
off-diagonal part
$\mathit{V}_\textrm{syst}{=}\mathit{V}_\textrm{unc}{+}\mathit{V}_\textrm{cor}$
(Table~\ref{tab:syst_combined}).  The $\mathit{V}_\textrm{unc}$ is
diagonal with $(\sigma_{\textrm{unc},\alpha})^2$.  The
$\mathit{V}_\textrm{cor}$ is the sum of a diagonal part with
$(\sigma_{\textrm{cor},\alpha})^2$ and an off-diagonal part with
$\rho_{\alpha\beta}\sigma_{\textrm{cor},\alpha}
\sigma_{\textrm{cor},\beta}$.  The correlation coefficient
$\rho_{\alpha\beta}$ is between two $D^0$ modes $\alpha$ and $\beta$.
The correlations among ${D^0}$-meson branching fractions are the PDG
values \cite{Amsler:2008zzb}; all others are assumed to be unity. 

\section{KINEMATIC DISTRIBUTIONS} 
\label{sec:dynamics} 

\noindent
This section presents the kinematic distributions
\cite{BMassConstraint}.  Sections~\ref{sec:3body}, \ref{sec:4body},
and \ref{sec:5body} give the plots for three-, four-, and five-body
modes, respectively.  Additional discussion is devoted to the
$M(p{\pi^-})$ feature in Sec.~\ref{sec:ppi}.

We briefly describe the background-subtraction and
efficiency-correction methods used to obtain the differential
branching fraction plots
(Figs.~\ref{fig:mass3body}--\ref{fig:mass5body}) as a function of
two-body invariant mass variables.  The differential branching
fraction, in bins $j$ of the plotted variable, is the ratio of the
number of signal events and the product of the correction factors as
given in Eq.~\ref{eq:formula}.  The quantity in the numerator is the
sum of the background-subtracted event weights for events in bin $j$;
the formulae are given below.  The efficiency-correction part of the
denominator is found for bin $j$ and is applied to each event weight.

The S-Plot method is used \cite{Pivk:2004ty} to find the event weight,
\begin{equation} 
    W(y_i) = 
    \frac{ 
        \rho_{\textrm{sig},\textrm{sig}}{P}_\textrm{sig}(y_i)+ 
        \rho_{\textrm{sig},\textrm{bgd}}{P}_\textrm{bgd}(y_i) 
    } 
    { 
        N_\textrm{sig}{P}_\textrm{sig}(y_i)+ 
        N_\textrm{bgd}{P}_\textrm{bgd}(y_i) 
    }, 
\end{equation} 

\noindent
where the $y_i$ is the pair of $\MES$ and $\DELTAE$ values for the
candidate in the $i^\textrm{th}$ event; the fit functions $P_\alpha$
were defined in Eq.~(\ref{eq:PDF}).  In general, the weight $W$ is
approximately $0$ for a background event and $1$ for a signal event.
The $\rho_{\textrm{sig,bgd}}$ quantifies the correlation between the
signal and the background yields,
\begin{equation} 
    (\rho_{\lambda,\lambda^\prime})^{-1} =  
    \sum_{i=1}^{N} 
    \frac{
        P_\lambda(y_i){P}_{\lambda^\prime}(y_i)
    }
    {
        \big( 
            N_\textrm{sig}{P}_\textrm{sig}(y_i){+} 
            N_\textrm{bgd}{P}_\textrm{bgd}(y_i) 
        \big)^2 
    }.
\end{equation} 

\subsection{\boldmath Three-body modes 
$\protect{B}{\rightarrow}{D^{(\!\ast\!)}}{p\pbar}$} 
\label{sec:3body}

\noindent
For the three-body modes, plots are given for Dalitz variables and
two-body invariant masses.

\begin{figure*}[tbp!] 
\centering 
\includegraphics[width=1\textwidth]{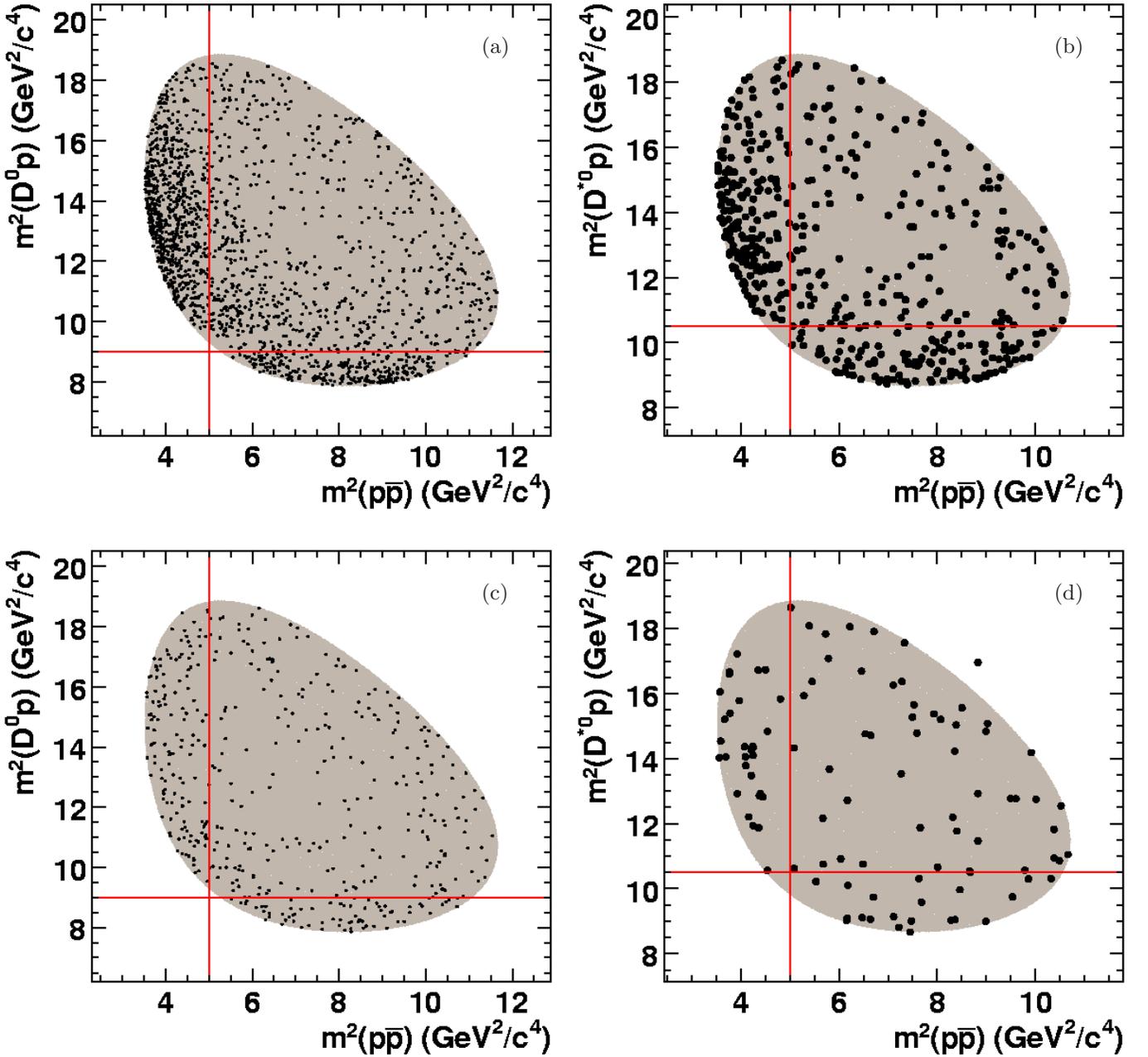} 
\caption{ 
    Dalitz plots $M^2({p\pbar})$ vs.~$M^2(D^{(\!\ast\!)0}{p})$ for the
    three-body modes.  Plots in the first column (a, c) correspond to
    ${\Bbar^0}{\rightarrow}{D^{0}}{p\pbar}$; the second column (b, d)
    ${\Bbar^0}{\rightarrow}{D^{\ast0}}{p\pbar}$.  Plots in the first
    row (a, b) are the events in the $\MES$-$\DELTAE$ signal box; the
    second row (c, d) the events in the $\MES$-sideband region
    normalized to the amount of background present in the respective
    plots in the first row.  In the first row, near-threshold
    enhancements are seen compared to the respective sideband plots in
    the second row.  The lines drawn at $M^2(p\pbar){=}5$,
    $M^2({D^0}{p}){=}9$, and $M^2({D^{\ast0}}{p}){=}10.5\gevccsq$ are
    visual aides to show that the enhancements are mostly
    non-overlapping.  The events are contained in the shaded contour
    representing the allowed kinematic region except for one outlier
    in (d), which failed the fit.  The points are made larger for the
    plots in the second column for better visibility.   
    }\label{fig:dalitz} 
\end{figure*} 

The Dalitz plots of $M^2(D^{(\!\ast\!)0}p)$ vs.~$M^2(p\pbar)$ for the
events in the $\MES$-$\DELTAE$ signal box are given
(Fig.~\ref{fig:dalitz}a, \ref{fig:dalitz}b).  The allowed kinematic
region is the shaded contour.  

The background events present in Figs.~\ref{fig:dalitz}a and
\ref{fig:dalitz}b are represented by Figs.~\ref{fig:dalitz}c and
\ref{fig:dalitz}d, respectively.  The latter plots show the events in
the $\MES$-sideband regions with their normalizations determined from
the background yield in the signal box.

\begin{figure*}[tbp!] 
\centering 
\includegraphics[width=1\textwidth]{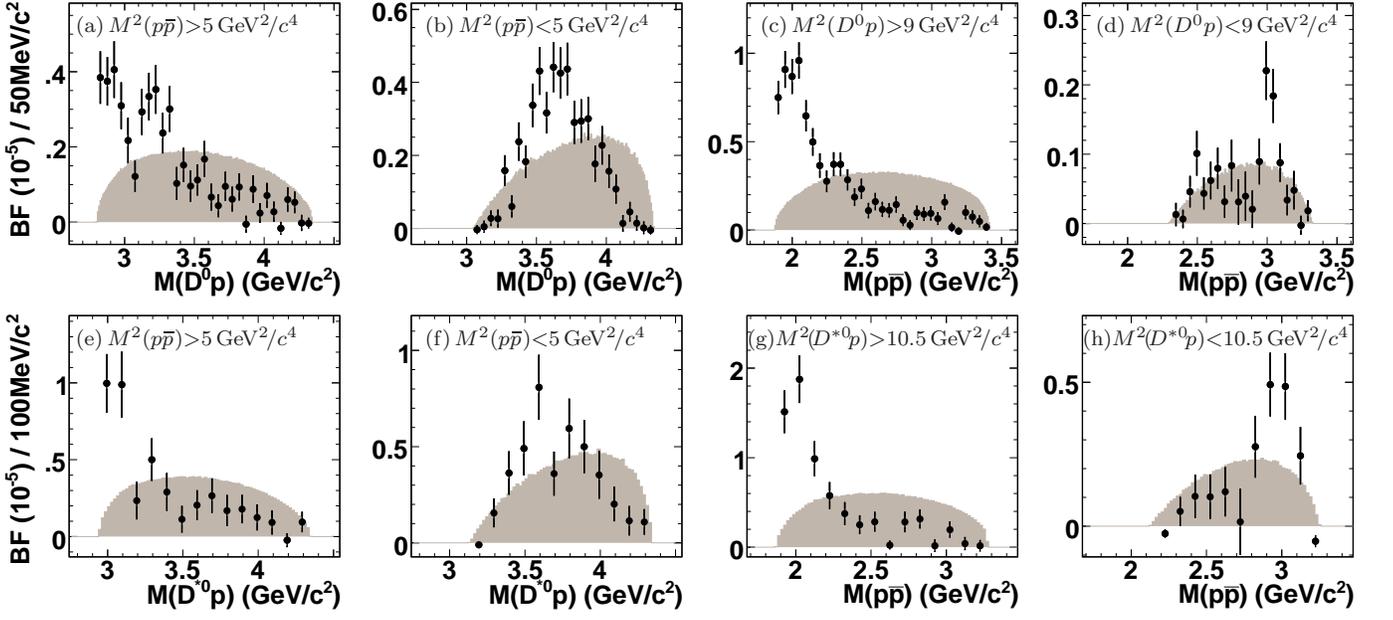} 
\caption{ 
    Differential branching fraction plots for the three-body $B$-meson
    modes:
    (a--d) ${\Bbar^0}{\rightarrow}{D^0}{p\pbar}$ and  
    (e--h) ${\Bbar^0}{\rightarrow}{D^{\ast0}}{p\pbar}$.  The captions
    give the various phase-space regions.  The shaded region
    represents the uniform phase-space model with its area normalized
    to the data.  The bin width for each row of plots is given on the
    left-most plot.
    }\label{fig:mass3body} 
\end{figure*}

The two-body invariant mass plots are given in
Fig.~\ref{fig:mass3body}.  Differential branching fractions are
plotted as a function of $M({D^{(\!\ast\!)0}}{p})$ and $M(p\pbar)$ for
events in different regions of the complementary variable.  The two
low-mass enhancements near threshold values in $M(D^{(\!\ast\!)0}p)$
and $M(p\pbar)$ correspond to the dense regions in the Dalitz plots.
The broad enhancement in Fig.~\ref{fig:mass3body}(d)
and~\ref{fig:mass3body}(h) does not have a substantial contribution
from $J/\psi$ decays due to its width and current experimental limits
on $\Bbar^0{\rightarrow}{D^0}{J/\psi}, {J/\psi}{\rightarrow}p\pbar$
\cite{Aubert:2005tr}.

In general, we observe a strong similarity between the shapes of the
corresponding distributions for
${\Bbar^0}{\rightarrow}{D^{0}}{p\pbar}$ and
${\Bbar^0}{\rightarrow}{D^{\ast0}}{p\pbar}$.

\subsection{\boldmath Four-body modes 
$\protect{B}{\rightarrow}{D^{(\!\ast\!)}}{p\pbar}\pi$} 
\label{sec:4body}

\begin{figure*}[p!] 
\centering 
\includegraphics[width=\textwidth]{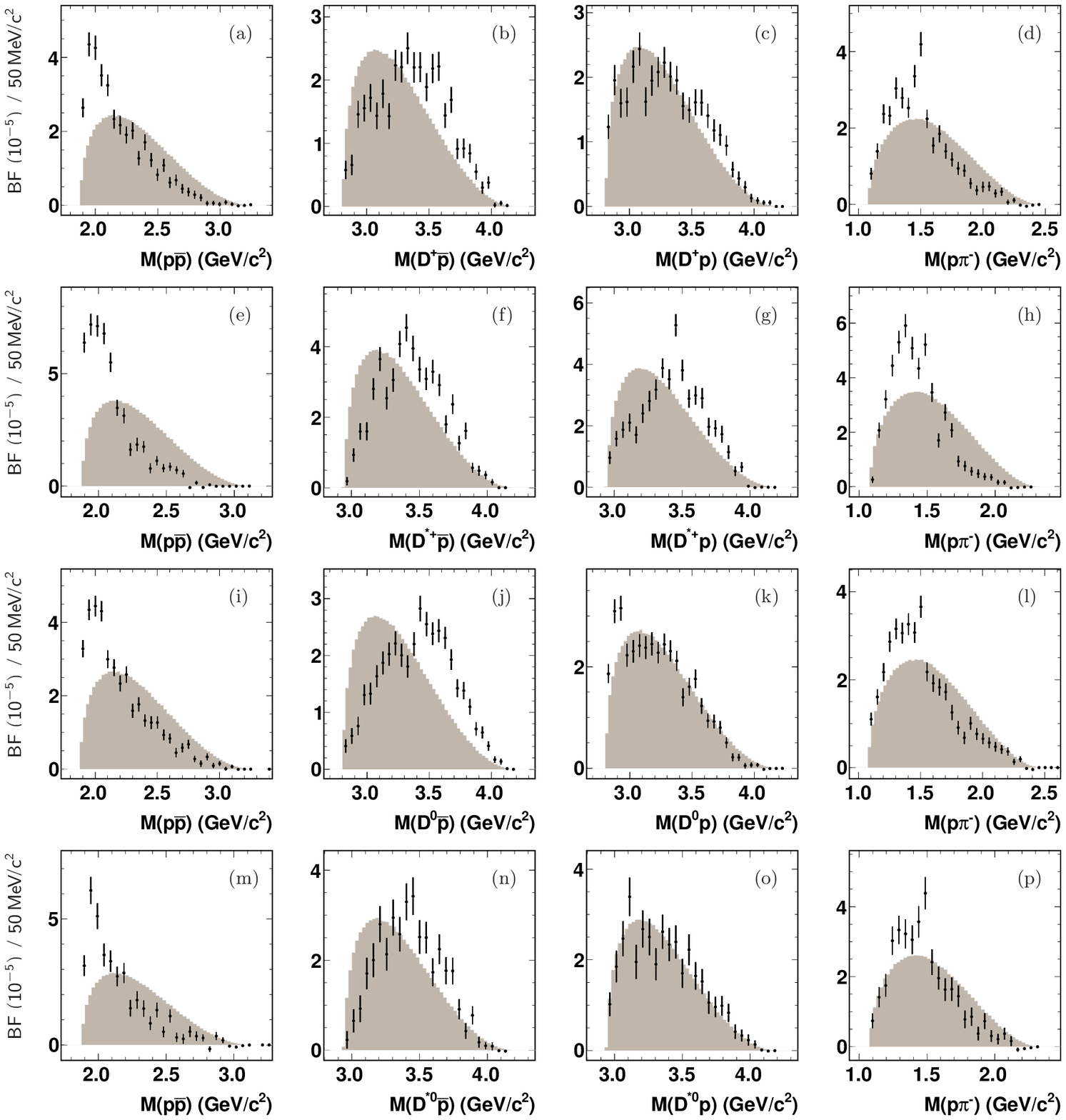} 
\caption{ 
    Differential branching fraction plots as 
    functions of $M({p\pbar})$, $M({D^{(\!\ast\!)}}{p})$, 
    $M({D^{(\!\ast\!)}}{\pbar})$, and $M(p{\pi^-})$ for the 
    four-body $B$-meson modes: 
    (a, b, c, d) ${\Bbar^0}{\rightarrow}{D^+}{p\pbar}{\pi^-}$, 
    (e, f, g, h) ${\Bbar^0}{\rightarrow}{D^{\ast+}}{p\pbar}{\pi^-}$, 
    (i, j, k, l) ${B^-}{\rightarrow}{D^0}{p\pbar}{\pi^-}$, and 
    (m, n, o, p) ${B^-}{\rightarrow}{D^{\ast0}}{p\pbar}{\pi^-}$, 
    respectively.  The shaded region represents the uniform
    phase-space model normalized to the data.  The possible presence
    of a narrow peak near $1.5\gevcc$ in plots in (d, h, j, p) are
    shown in detail in Figs.~\ref{fig:massoppsign}a,
    \ref{fig:massoppsign}b, \ref{fig:massoppsign}c, and
    \ref{fig:massoppsign}d, respectively, and discussed in
    Sec.~\ref{sec:ppi}.  The bin width for each row of plots is given on
    the left.
}\label{fig:mass4body} 
\end{figure*}

\noindent
For the four-body modes, plots are given for two-body invariant masses
in Fig.~\ref{fig:mass4body}.  Differential branching fractions are
plotted as a function of $M({p\pbar})$, $M({D^{(\!\ast\!)}}{p})$,
$M({D^{(\!\ast\!)}}{\pbar})$, and $M(p{\pi^-})$.

The two-body invariant-mass distributions show a number of features.
The $M({p\pbar})$ distributions show a threshold enhancement with
respect to the expectations from the uniform phase-space decay model
(Figs.~\ref{fig:mass4body}a, \ref{fig:mass4body}e,
\ref{fig:mass4body}i, \ref{fig:mass4body}m).  The
$M({D^{(\!\ast\!)}}{\pbar})$ distributions show no indication of a
penta-quark resonance at $3.1\gevcc$ \cite{Aktas:2004qf}
(Figs.~\ref{fig:mass4body}b, \ref{fig:mass4body}f,
\ref{fig:mass4body}j, \ref{fig:mass4body}n).  The
$M({D^{(\!\ast\!)}}{p})$ distribution in one of the modes
(Fig.~\ref{fig:mass4body}k) suggests a threshold enhancement, as was
observed in the three-body modes, but the distributions in the other
modes show no such features (Figs.~\ref{fig:mass4body}c,
\ref{fig:mass4body}g, \ref{fig:mass4body}o).  The $M(p{\pi^-})$
distribution in one of the modes (Fig.~\ref{fig:mass4body}d) shows a
narrow structure near $1.5\gevcc$, but it is less prominent in the
distributions of the other modes (Figs.~\ref{fig:mass4body}h,
\ref{fig:mass4body}l, \ref{fig:mass4body}p).  

The peak near $1.5\gevcc$ does not correspond to a known state.  The
peak is discussed in detail in Sec.~\ref{sec:ppi}.

\subsection{\boldmath Five-body modes 
$\protect{B}{\rightarrow}{D^{(\!\ast\!)}}{p\pbar}\pi\pi$} 
\label{sec:5body}

\begin{figure*}[p!] 
\centering 
\includegraphics[width=\textwidth]{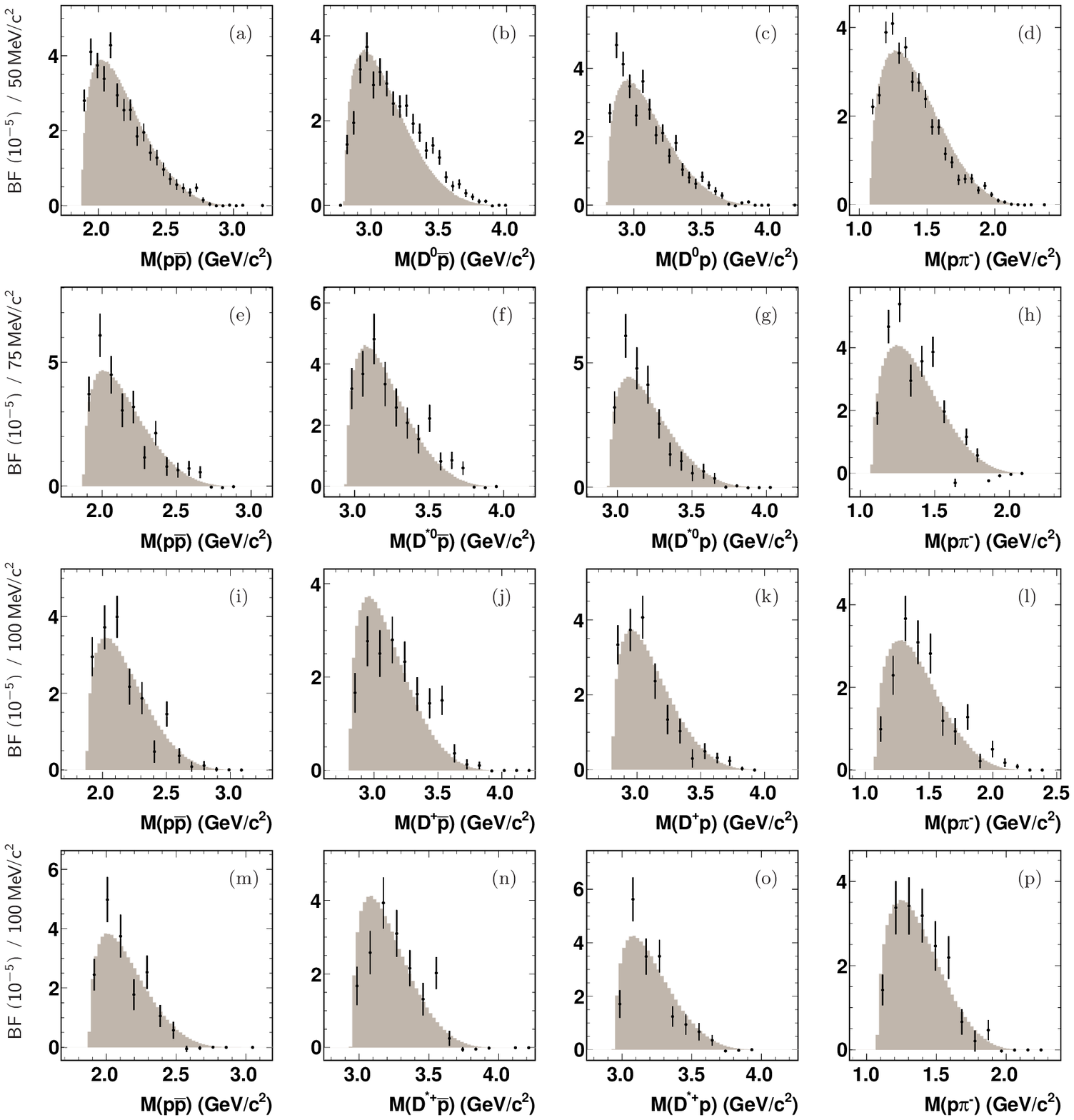} 
\caption{ 
    Differential branching fraction plots as 
    functions of $M({p\pbar})$, $M({D^{(\!\ast\!)}}{p})$, 
    $M({D^{(\!\ast\!)}}{\pbar})$, and $M(p{\pi^-})$ for five-body 
    $B$-meson modes: 
    (a, b, c, d) ${\Bbar^0}{\rightarrow}{D^0}{p\pbar}{\pi^-}{\pi^+}$, 
    (e, f, g, h) ${\Bbar^0}{\rightarrow}{D^{\ast0}}{p\pbar}{\pi^-}{\pi^+}$, 
    (i, j, k, l) ${B^-}{\rightarrow}{D^+}{p\pbar}{\pi^-}{\pi^-}$, and 
    (m, n, o, p) ${B^-}{\rightarrow}{D^{\ast+}}{p\pbar}{\pi^-}{\pi^-}$,
    respectively.  The shaded region represents the uniform
    phase-space model normalized to the data.  Each $B$-meson
    candidate for the plots in (d, h) contributes two entries for both
    $p\pi^-$ and $\pbar\pi^+$ combinations, so they are scaled
    accordingly.  The bin width for each row of plots is given on the
    left.
}\label{fig:mass5body} 
\end{figure*}

\noindent
For the five-body modes, plots are given for two-body invariant masses
in Fig.~\ref{fig:mass5body}.  Branching fractions are plotted as a
function of $M({p\pbar})$, $M({D^{(\!\ast\!)}}{p})$,
$M({D^{(\!\ast\!)}}{\pbar})$, and $M(p{\pi^-})$.

In contrast to the distributions for the three- and four-body modes,
the five-body distributions are generally more consistent with the
expectations from the uniform phase-space decay model. 

A notable absence, again, is the signal of a penta-quark resonance at
$3.1\gevcc$ \cite{Aktas:2004qf} (Figs.~\ref{fig:mass5body}b,
\ref{fig:mass5body}f, \ref{fig:mass5body}j, \ref{fig:mass5body}n). 

\subsection{\boldmath Narrow $M(p{\pi^-})$ peak at $1.5\gevcc$} 
\label{sec:ppi} 

\noindent
The narrow peak in the $M(p{\pi^-})$ \cite{fn:sign} distribution at
$1.5\gevcc$, which we refer to as $X$, is discussed in this section.

\begin{figure*}[tbp!] 
\centering 
\includegraphics[width=0.893\textwidth]{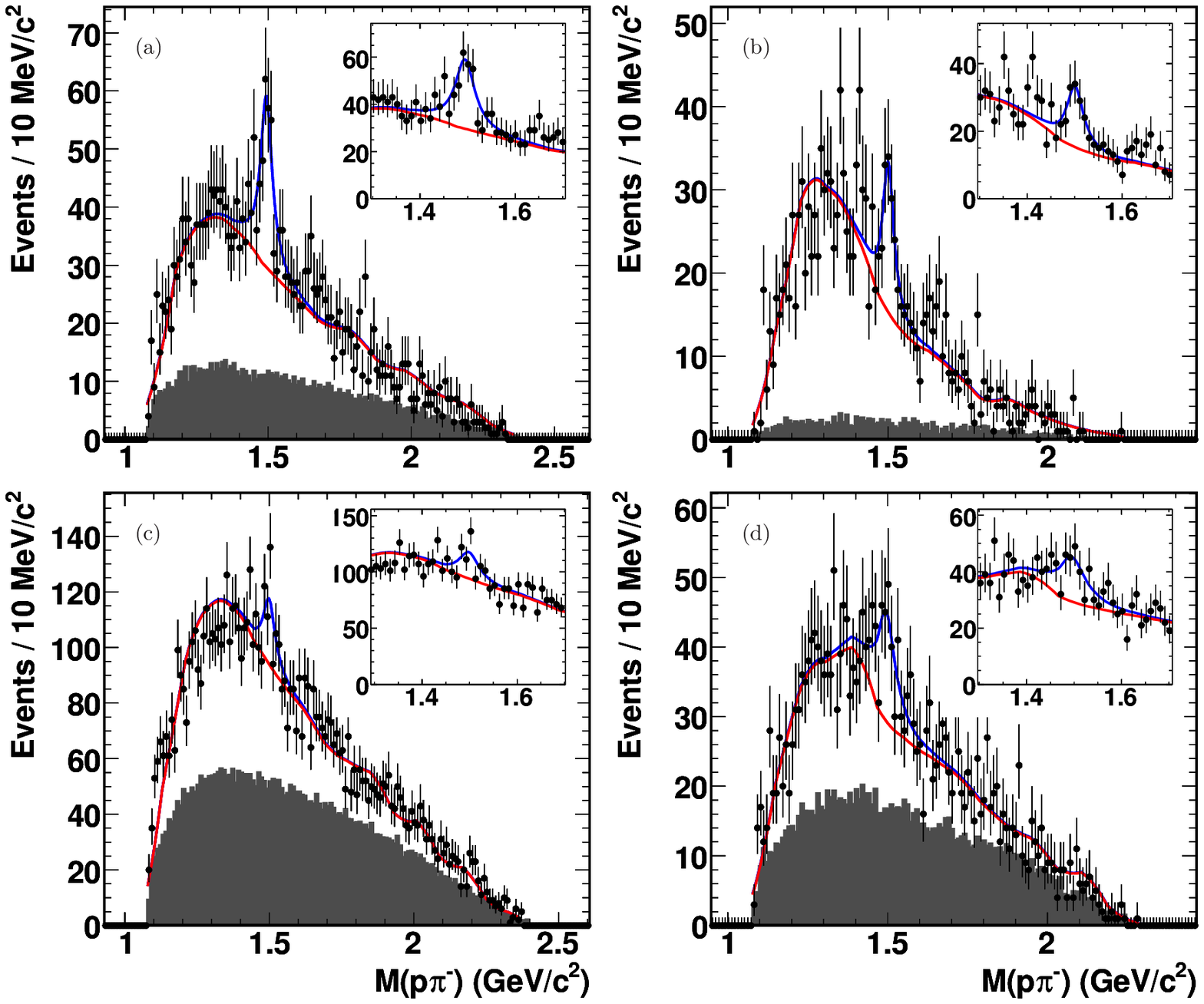} 
\caption{ 
    Fits of the opposite-sign  $M(p{\pi^-})$ distribution for
    (a) ${\Bbar^0}{\rightarrow}{D^+}{p\pbar}{\pi^-}$, 
    (b) ${\Bbar^0}{\rightarrow}{D^{\ast+}}{p\pbar}{\pi^-}$,  
    (c) ${B^-}{\rightarrow}{D^0}{p\pbar}{\pi^-}$, and  
    (d) ${B^-}{\rightarrow}{D^{\ast0}}{p\pbar}{\pi^-}$ for events
    in the signal box of $\MES$-$\DELTAE$.  The top curve is the sum
    of $P_\textrm{sig}$ and $P_\textrm{bgd}$ while the bottom curve is
    $P_\textrm{bgd}$.  The $P_\textrm{bgd}$ is from the corresponding
    plot in Fig.~\ref{fig:masssamesign}.  The shaded histograms are
    scaled $\MES$ sidebands.  A small inset plot is a close-up of the
    region around $1.5\gevcc$; its bin width is the same as in the
    larger plot.   
    }\label{fig:massoppsign} 
\end{figure*}

The opposite-sign $M(p\pi^-)$ distributions corresponding to 
Figs.~\ref{fig:mass4body}d, \ref{fig:mass4body}h,
\ref{fig:mass4body}l, \ref{fig:mass4body}p are shown in more detail in
Fig.~\ref{fig:massoppsign}.  In the detailed plots, the $x$-axis bin
width is smaller at $10\mevcc$ and the $y$-axis is the
unweighted-uncorrected number of events.  The events from the
$\MES$-sideband region is superimposed with its normalization
determined from the background yield in the $\MES$-$\DELTAE$ signal
box.

\begin{figure*}[tbp!] 
\centering 
\includegraphics[width=0.893\textwidth]{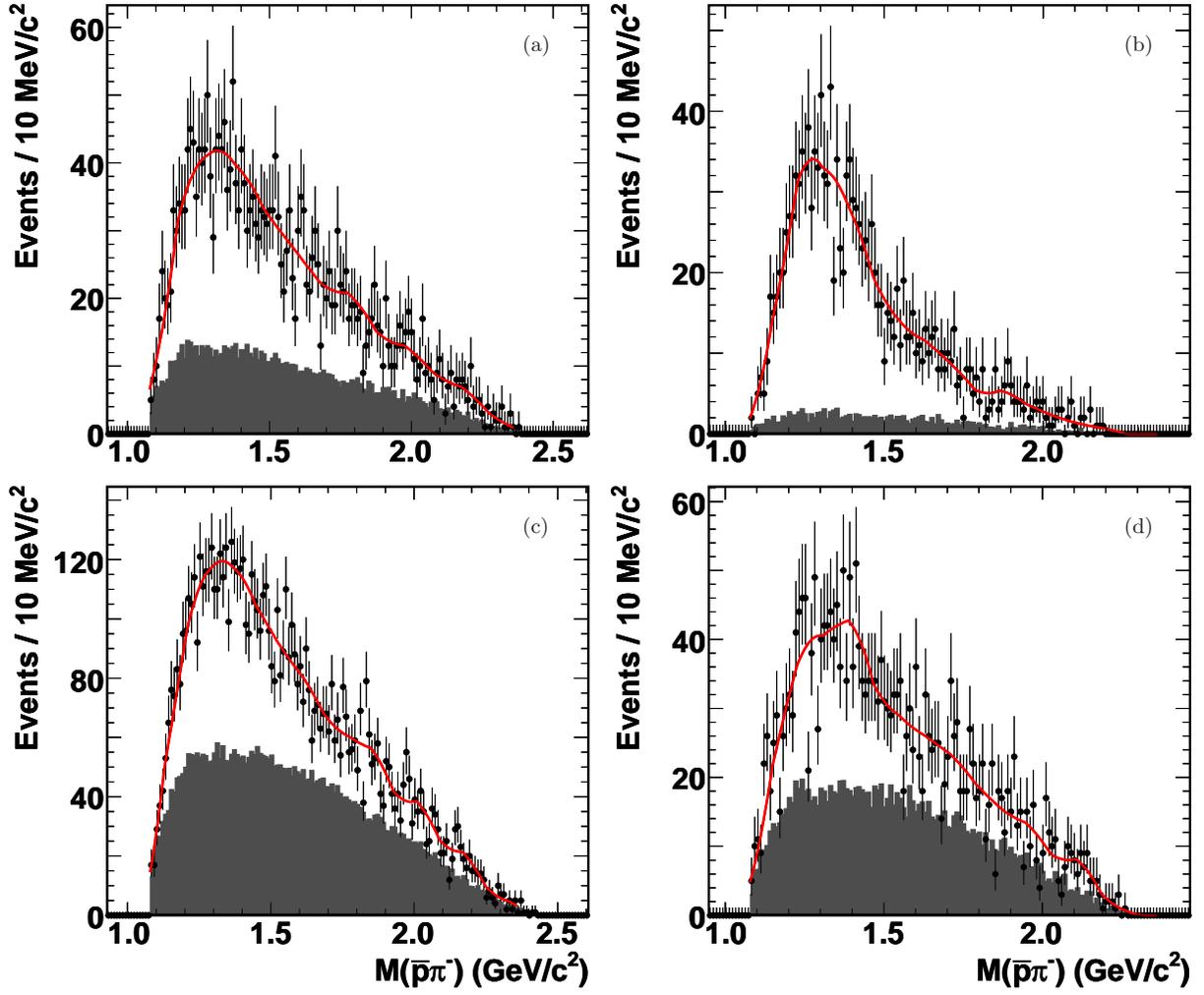} 
\caption{ 
    Fits of the same-sign  $M(\pbar{\pi^-})$ distribution for
    (a) ${\Bbar^0}{\rightarrow}{D^+}{p\pbar}{\pi^-}$,  
    (b) ${\Bbar^0}{\rightarrow}{D^{\ast+}}{p\pbar}{\pi^-}$,  
    (c) ${B^-}{\rightarrow}{D^0}{p\pbar}{\pi^-}$, and  
    (d) ${B^-}{\rightarrow}{D^{\ast0}}{p\pbar}{\pi^-}$ for events
    in the signal box of $\MES$-$\DELTAE$.  The curve is the smoothed
    histogram that is used in the corresponding plot in
    Fig.~\ref{fig:massoppsign} as $P_\textrm{bgd}$.  The shaded
    histograms are scaled $\MES$ sidebands.  
    }\label{fig:masssamesign} 
\end{figure*}

In order to measure the properties of the peak, the fit formalism of
Eq.~(\ref{eq:PDF}) is used.  The signal component $P_{\textrm{sig}}$
is assumed to be a Breit-Wigner line shape.  The background component
$P_{\textrm{bgd}}$ is taken from the same-sign $M(\pbar{\pi^-})$
distribution.  The distribution for the
$\Bbar^0{\rightarrow}{D^+}{p}{\pbar}{\pi^-}$ mode is relatively smooth
(Fig.~\ref{fig:masssamesign}a), and it describes the rise and fall of
the opposite-sign distribution well (Fig.~\ref{fig:massoppsign}a),
whereas the same-sign distributions in the other modes show a more
rapidly falling behavior around $1.5\gevcc$
(Figs.~\ref{fig:masssamesign}b--\ref{fig:masssamesign}d).

We note, however, that the use of the shape for $P_{\textrm{bgd}}$ has
limitations.  Since the formation of the $p$ or $\pbar$ is not
necessarily symmetric with respect to the $\pi^-$ in these decays, the
same-sign $M(\pbar\pi^-)$ combination may not predict the true shape
for the non-resonant component in the opposite-sign $M(p\pi^-)$
distribution.  As a consequence, we cannot precisely quantify the
systematic uncertainty associated with the lack of knowledge of the
true background shape. 

For the two neutral $B$ modes, the fits of the opposite-sign
distributions describe the entire kinematic range well
(Figs.~\ref{fig:massoppsign}a, \ref{fig:massoppsign}b).  We note a
small excess of events above $1.65\gevcc$ with respect to
$P_{\textrm{bgd}}$, but no peak component is included in the fit at
this mass.  The fitted $X$ mass is $1494.4{\pm}4.1\mevcc$ and
$1500.8{\pm}4.4\mevcc$, where the uncertainties are statistical, for
${\Bbar^0}{\rightarrow}{D^{\ast+}}{p\pbar}{\pi^-}$ and
${\Bbar^0}{\rightarrow}{D^{+}}{p\pbar}{\pi^-}$, respectively.  We
measure the full widths to be $51{\pm}18\mev$ and $43{\pm}17\mev$,
respectively.  The widths are significantly wider than detector
resolution, which is less than $4\mev$ for a simulated
$X{\rightarrow}p{\pi^-}$ decay with a mass of $1.5\gevcc$ and
negligible width. 

In contrast to the neutral $B$ modes, the opposite-sign distributions
for the two charged $B$ modes exhibit a less peaking behavior at
$1.5\gevcc$.  As a result, the parameter for the width in the
${B^-}{\rightarrow}{D^0}{p\pbar}{\pi^-}$ mode is fixed to the value
found in the ${\Bbar^0}{\rightarrow}{D^+}{p\pbar}{\pi^-}$ mode; the
results of this fit are not used in the average.

\begin{figure*}[tbp!] 
\centering 
\includegraphics[width=0.893\textwidth]{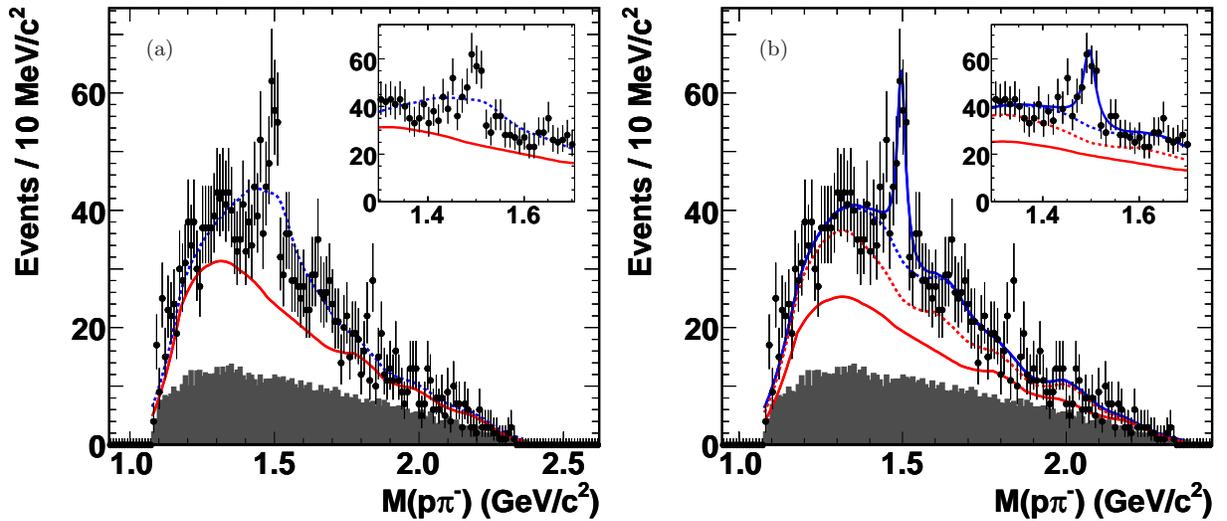} 
\caption{ 
    Alternate fits of the opposite-sign $M(p{\pi^-})$ distribution for
    ${\Bbar^0}{\rightarrow}{D^+}{p\pbar}{\pi^-}$ with 
    (a) various $N^\ast$ resonances and 
    (b) an additional $P_\textrm{bgd}$ obtained from the 
    ${\Bbar^0}{\rightarrow}{D^{\ast+}}{p\pbar}{\pi^-}$ sample. 
    The shaded histograms are the scaled $\MES$-$\DELTAE$ sidebands.
    A small inset plot is a close-up of the region around $1.5\gevcc$;
    its bin width is the same as the larger plot.   
    }\label{fig:massoppsignalt} 
\end{figure*}

The known nucleon resonances $N^\ast$ with the masses $1440$, $1520$,
$1535$, and $1650\mevcc$ are used in an attempt to describe the $X$.
The distribution is fit with the $N^\ast$ fit function components each
parameterized as a Breit-Wigner line shape.  The normalization for each
component is allowed to vary independently.  However, the fit does not
describe the peak because the $X$ is much narrower than any of the
$N^\ast$ resonances (Fig.~\ref{fig:massoppsignalt}a).

The overall significance of the $X$ is difficult to measure, due to
our lack of knowledge of the true background shape, as discussed
earlier, as well as further statistical issues.  We caution that the
$X$ analysis is not blind, the parameters are not chosen a priori, and
the distribution under the no-$X$ hypothesis may be only approximately
normal. Furthermore, even under the normal assumption, the presence of
the mass and width nuisance parameters under the alternative
hypothesis means that the distributions of the $S$ statistic is not
likely to be pure $\chi^2$. 

We provide a measure of the statistical significance
$S{=}\sqrt{2(\ln{L}_1{-}\ln{L}_0)}$ of the $X$ in the two neutral $B$
modes, where $L_1$ is the likelihood value with $P_{\textrm{sig}}$ and
$L_0$ is without $P_{\textrm{sig}}$.  The value is $S{=}8.6$ for
${\Bbar^0}{\rightarrow}{D^+}{p\pbar}{\pi^-}$ and $S{=}6.9$ for
${\Bbar^0}{\rightarrow}{D^{\ast+}}{p\pbar}{\pi^-}$.

The systematic uncertainties are mainly due to the $P_{\textrm{bgd}}$.
We fit using an alternate fit function by adding a component derived
from the same-sign distribution of a different mode
(Fig.~\ref{fig:massoppsignalt}b).  The result is a mass shift of
$0.8\mevcc$ and a full width change of $4\mev$.  An additional
contribution of $0.5\mevcc$ is added for the mass measurement due to
the absolute uncertainty of the magnetic field and the amount of
detector material \cite{Aubert:2005gt}. 

In summary, the unknown structure $X$ can be characterized by a
Breit-Wigner line shape:
\begin{eqnarray} 
M(X)      &=& 1497.4\pm 3.0\pm 0.9\mevcc \\ 
\Gamma(X) &=& \phantom{00.0}47\pm 12\phantom{.}\pm 4\phantom{.0}\mev, \nonumber 
\end{eqnarray} 

\noindent
where the uncertainties are statistical and systematic, respectively. 

\section{CONCLUSIONS} 
\label{sec:conclusions} 

\noindent 
We have presented a study of ten baryonic $B$-meson decay modes of the
form $B{\rightarrow}{D^{(\!\ast\!)}}{p\pbar}(\!\pi\!)(\!\pi\!)$  using
a data sample of $455{\times}10^6$ $B\Bbar$ pairs.  Significant
signals are observed (Table \ref{tab:bf_chain}).  Six of the
modes---${B^-}{\rightarrow}{D^0}{p\pbar}{\pi^-}$,
${B^-}{\rightarrow}{D^{\ast0}}{p\pbar}{\pi^-}$,
${\Bbar^0}{\rightarrow}{D^0}{p\pbar}{\pi^-}{\pi^+}$,
${\Bbar^0}{\rightarrow}{D^{\ast0}}{p\pbar}{\pi^-}{\pi^+}$,
${B^-}{\rightarrow}{D^+}{p\pbar}{\pi^-}{\pi^-}$, and
${B^-}{\rightarrow}{D^{\ast+}}{p\pbar}{\pi^-}{\pi^-}$---are observed
for the first time (Figs.~\ref{fig:mes4body}e--\ref{fig:mes4body}g,
\ref{fig:mes4body}h--\ref{fig:mes4body}j, \ref{fig:mes5body}a,
\ref{fig:mes5body}d, \ref{fig:mes5body}g,
\ref{fig:mes5body}h--\ref{fig:mes5body}j, respectively).

The $B$-meson branching fraction measurements range from
$0.97{\times}10^{-4}$ to $4.55{\times}10^{-4}$ with the hierarchy
$\mathcal{B}_\textrm{\,3-body}{<}\mathcal{B}_\textrm{\,5-body}{<}\mathcal{B}_\textrm{\,4-body}$
(Table \ref{tab:bf_mode}).  These results supersede the previous
{\Babar} publication of ${\Bbar^0}{\rightarrow}{D^0}{p\pbar}$,
$D^{\ast0}{p\pbar}$, ${D^+}{p\pbar}{\pi^-}$, and
$D^{\ast+}{p\pbar}{\pi^-}$ \cite{Aubert:2006qx}.  The branching
fractions related by changes in the charge or the spin of the $D$
meson are found to be similar (Table \ref{tab:bf_ratio}).

The kinematic distributions show a number of notable features.  For
the three-body modes, threshold enhancements are present in
$M({p\pbar})$ and $M({D^{(\!\ast\!)0}}{p})$ (Figs.~\ref{fig:dalitz},
\ref{fig:mass3body}).  For the four-body modes, a threshold
enhancement is observed in $M(p{\pbar})$ and a narrow peak is seen in
$M(p{\pi^-})$ (Fig.~\ref{fig:mass4body}).  For the five-body modes, in
contrast to the other modes, the distributions are similar to the
expectations from the uniform phase-space decay model
(Fig.~\ref{fig:mass5body}).

The $M(p{\pi^-})$ distributions in the neutral $B$-meson decay mode
show the most prominent peak near $1.5\gevcc$.  We obtained a mass of
$1497.4{\pm}3.0{\pm}0.9\mevcc$ and a full width of
$47{\pm}12{\pm}4\mev$, where the first uncertainties are statistical
and the second are systematic, respectively
(Figs.~\ref{fig:massoppsign}--\ref{fig:massoppsignalt}).  Determining
the significance and interpreting the origin of the peak are
complicated by the fact that the background fit function is
parameterized by the distribution from the same-sign charge
combinations ${\pbar}{\pi^-}$, a procedure which may not provide the
true background shape. 

Despite the relatively small branching fractions for these modes of
order $10^{-4}$, with product branching fractions of order $10^{-5}$
to $10^{-6}$ (including the $D$ and $D^{\ast}$ modes), the large size
of the {\Babar} data sample allowed us to observe signals containing
hundreds of events in many of the modes.  We are, therefore, able to
probe their kinematic distributions that reflect the complex dynamics
of the multi-body final states.

\section{ACKNOWLEDGMENTS}
\label{sec:Acknowledgments} 

\noindent
We are grateful for the extraordinary contributions of our PEP-II
colleagues in achieving the excellent luminosity and machine
conditions that have made this work possible.  The success of this
project also relies critically on the expertise and dedication of the
computing organizations that support {\Babar}.  The collaborating
institutions wish to thank SLAC for its support and the kind
hospitality extended to them.  This work is supported by the US
Department of Energy and National Science Foundation, the Natural
Sciences and Engineering Research Council (Canada), the Commissariat
\`a l'Energie Atomique and Institut National de Physique Nucl\'eaire
et de Physique des Particules (France), the Bundesministerium f\"ur
Bildung und Forschung and Deutsche Forschungsgemeinschaft (Germany),
the Istituto Nazionale di Fisica Nucleare (Italy), the Foundation for
Fundamental Research on Matter (The Netherlands), the Research Council
of Norway, the Ministry of Education and Science of the Russian
Federation, Ministerio de Ciencia e Innovaci\'on (Spain), and the
Science and Technology Facilities Council (United Kingdom).
Individuals have received support from the Marie-Curie IEF program
(European Union), the A. P. Sloan Foundation (USA) and the Binational
Science Foundation (USA-Israel).

\end{document}

%% file: authors_jun2010_bad2210.tex
%% author list as of 01-Jun-2010 (440 authors)
%
\author{P.~del~Amo~Sanchez}
\author{J.~P.~Lees}
\author{V.~Poireau}
\author{E.~Prencipe}
\author{V.~Tisserand}
\affiliation{Laboratoire d'Annecy-le-Vieux de Physique des Particules (LAPP), Universit\'e de Savoie, CNRS/IN2P3,  F-74941 Annecy-Le-Vieux, France}
\author{J.~Garra~Tico}
\author{E.~Grauges}
\affiliation{Universitat de Barcelona, Facultat de Fisica, Departament ECM, E-08028 Barcelona, Spain }
\author{M.~Martinelli$^{ab}$}
\author{A.~Palano$^{ab}$ }
\author{M.~Pappagallo$^{ab}$ }
\affiliation{INFN Sezione di Bari$^{a}$; Dipartimento di Fisica, Universit\`a di Bari$^{b}$, I-70126 Bari, Italy }
% ...
\collaboration{The \babar\ Collaboration}
\noaffiliation

%% file: paper.bbl
\begin{thebibliography}{99}
\bibitem{fn:charge} 
    We use the convention that charge conjugation of particles and their
    decays is implied unless otherwise specified.
\bibitem{fn:dmass} 
    We follow the notation where $M(D)$ is the invariant mass of the
    reconstructed daughters of the $D$-meson candidate.
\bibitem{fn:sign} 
    We refer to both $p\pi^-$ and $\pbar\pi^+$ as opposite sign and to
    both $\pbar\pi^-$ and $p\pi^+$ as same sign.
\end{thebibliography}
